%% file: main.tex
\documentclass[a4paper,11pt]{article}

\usepackage{graphicx} 
\usepackage{titlesec}
\usepackage{subfiles}
\usepackage{adjustbox}
\usepackage{booktabs}
\usepackage{multirow}
\usepackage{multicol}
\usepackage{placeins}
\usepackage{booktabs}
\usepackage{hyperref}
\usepackage{subfig}
\usepackage{subcaption}
\usepackage{appendix}
\hypersetup{colorlinks=true,linkcolor=blue}

\usepackage{geometry}
\usepackage{indentfirst}
\usepackage{amsmath, amssymb, amsfonts}

\usepackage[
backend=biber,
style=numeric,
sorting=none
]{biblatex}   

\DeclareFieldFormat[article]{journaltitle}{#1\isdot}
\DefineBibliographyStrings{english}{%
    in = {},
}

\newcommand{\beq}{\begin{equation}}
\newcommand{\eeq}{\end{equation}}
\newcommand{\ba}{\begin{array}}
\newcommand{\ea}{\end{array}}
\newcommand{\beqa}{\begin{eqnarray}}
\newcommand{\eeqa}{\end{eqnarray}}
\newcommand{\beqs}{\begin{subequations}}
\newcommand{\eeqs}{\end{subequations}}

\renewcommand{\rm}{\textnormal}

\def\leqq{\leqslant}
\def\geqq{\geqslant}
\def\({\left(}
\def\){\right)}
\def\[{\left[}
\def\]{\right]}

\def\to{\rightarrow}

\def\leqq{\leqslant}
\def\geqq{\geqslant}
\def\hs{\hspace*{0.3mm}}

\def\hsm{\hspace*{-0.3mm}}

\allowdisplaybreaks

\title{\bf Probing Neutral Triple Gauge Couplings\\ 
via $\boldsymbol{Z\gamma\,(\ell^+\ell^-\gamma)}$ Production
at $\boldsymbol{e^+e^-}$ Colliders}
\author{ 
\textbf{\large Danning Liu$^{1,2}$, Rui-Qing Xiao$^{1,2,3}$, 
Shu Li$^{1,2,4}$,}\\
\vspace{5mm}
\textbf{\large John Ellis$^{3,5,1}$, Hong-Jian He$^{1,2,4}$, 
Rui Yuan$^{1,2}$} 
}
\date{
$^1$\,{Tsung-Dao Lee Institute, Shanghai Jiao Tong University,   
Shanghai, China} \\
$^2$\,{School of Physics and Astronomy, 
Key Laboratory for Particle Astrophysics\\
and Cosmology, 
Shanghai Key Laboratory for Particle Physics and Cosmology, \\
Shanghai Jiao Tong University, Shanghai, China} \\
$^3$\,{Department of Physics, King's College London, Strand, London WC2R 2LS, UK}    \\
$^4$\,{Center for High Energy Physics, Peking University,  
Beijing, China}     \\
$^5$\,{Theoretical Physics Department, CERN, 
CH-1211 Geneva 23, Switzerland}  
\\[5mm]
{\tt \ danningliu@sjtu.edu.cn, xiaoruiqing@sjtu.edu.cn, shuli@sjtu.edu.cn, 
\\
john.ellis@cern.ch, hjhe@sjtu.edu.cn, yuanrui@sjtu.edu.cn}
}

\begin{document}
\thispagestyle{empty}

\maketitle

\begin{abstract}
Neutral triple gauge couplings (nTGCs) are absent in the Standard Model (SM) and at the
dimension-6 level in the Standard Model Effective Field Theory (SMEFT), arising first from dimension-8 operators.\ 
As such, they provide a unique window for probing new physics beyond the SM.\   
These dimension-8 operators can be mapped to nTGC form factors whose structure is consistent
with the spontaneously-broken electroweak gauge symmetry of the SM.\ 
In this work, we study the probes of nTGCs in the reaction 
$e^+e^-\!\!\to\! Z\gamma\,$ with $Z\!\!\to\! \ell^+\ell^-\,(\ell\!=\!e,\mu)$ at an $e^+ e^-$ collider.\  
We perform a detector-level
simulation and analysis of this reaction  
at the Circular Electron Positron Collider (CEPC) 
with collision energy $\sqrt{s} \!=\! 240$\,GeV
and an integrated luminosity of 20\,ab$^{-1}$.\ 
We present the sensitivity limits 
on probing the new physics scales of dimension-8 nTGC operators 
via measurements of the corresponding nTGC form factors. 
\\[4mm]
Frontiers of Physics (2024), in Press [arXiv:2404.15937].\\[2mm]
KCL-PH-TH/2024-18, CERN-TH-2024-046
\end{abstract}
\thispagestyle{empty}
\newpage

\clearpage

\tableofcontents

\section{Introduction}
\subfile{introduction.tex}
\label{sec:intro}
\label{sec:1}

\section{Theoretical Framework, Simulation and Analysis}
\subfile{Combined_Chapter2}
\label{sec:simuandana}
\label{sec:2}

\section{Systematics}
\subfile{Systematics.tex}
\label{sec:systematics}
\label{sec:3}

\section{Results and Discussions}
\subfile{Results.tex}
\label{sec:results}
\label{sec:4}
\FloatBarrier

\section{Conclusions}
\subfile{Conclusions.tex}

\label{sec:5}

\vspace*{10mm}
\noindent
{\large\bf Acknowledgments}
\\[1mm]
The work of J.E. was supported in part by the United Kingdom STFC Grant ST/T000759/1.\ 
The work of HJH and RQX was supported in part by the NSFC Grants 12175136 and 11835005.\ 
RQX has also been supported by an International Postdoctoral Exchange Fellowship.\ 
We thank Gang Li, Yulei Zhang and Xuliang Zhu for discussions of the CEPC detector configuration.

\newpage
\appendix 
\renewcommand{\thesubsection}{\thesection.\arabic{subsection}}
\noindent
{\huge\bf Appendix}
\addcontentsline{toc}{section}{Appendix}
\subfile{Appendix.tex}
\label{sec:appendix}

\newpage 
\printbibliography[
heading=bibintoc,
title={References}
]

\end{document}

%% file: Introduction.tex

The Standard Model Effective Field Theory (SMEFT)~\cite{Buchmuller:1985jz} is a powerful framework 
for studying model-independently possible new physics beyond the Standard Model (SM).\ 
The SMEFT includes only the known elementary particles, which are assumed to have the SM quantum numbers 
and thus have the interactions with mass-dimension $d\!\leqq\! 4\hs$ that are predicted by the SM, but the SMEFT also
includes additional effective interactions with mass-dimensions $d \!>\! 4\hs$.\ 
Such higher-dimensional interactions could arise from new physics at energy scales beyond the electroweak scale due to possible
exchanges of new massive particles and/or novel strong dynamics.\ 
The SMEFT interactions with dimension\,5 may be relevant 
for neutrino physics, whereas collider experiments are 
generally sensitive to SMEFT interactions with 
even dimensions $d \hsm\geqq\hsm 6\hs$.\ 
Probing the effects of SMEFT operators may either constrain the possible high-scale 
new physics dynamics or provide hints to its possible nature, 
without assuming the ultraviolet (UV) origin or   
making any assumptions about its form.

There is an extensive theoretical literature classifying the SMEFT operators of dimension 6~\cite{Grzadkowski:2010es,Giudice:2007fh} and above~\cite{Murphy:2020rsh,Li:2020gnx}, and a growing number of phenomenological and experimental papers analyzing the constraints on their possible coefficients that are imposed by current data from the LHC and elsewhere. Most of these analyses have had operators with $d= 6$ as their primary focus, often working to linear order in the SMEFT operator coefficients, i.e., quadratically in the new physics scale, an approximation that takes into account their interference with SM interactions~\cite{Pomarol:2013zra,Berthier:2015oma,Berthier:2015gja,Biekotter:2018ohn,Ellis:2020unq}. To date there is no significant indication that any $d=6$ SMEFT operator has a non-zero coefficient, but future colliders will provide much greater precision in SMEFT probes~\cite{He:2015spf,Ge:2016zro,Ge:2016tmm,deBlas:2019rxi}.

A complete analysis of the phenomenology of dimension-6 operators should include their quadratic effects on event rates, which depend quartically on the new physics scale. At this level one should in general consider the effects of linear interference between dimension-8 SMEFT operators and SM amplitudes, which also depend quartically on the new physics scale, and there is a growing literature of analyses that take these into account~\cite{Corbett:2021eux,Corbett:2023qtg,Brivio:2019ius,Durieux:2019rbz,Ethier:2021bye}. Complementing these studies, it is interesting to consider processes that have no dimension-6 operator contributions, to which dimension-8 operators make the leading SMEFT contributions. These processes include quartic neutral vector-boson interactions and also neutral triple gauge couplings (nTGCs), where the nTGCs are the object of the present study.

Neutral triple gauge couplings are absent in the SM and at the level of dimension-6 operators in the SMEFT, arising first at the level of dimension-8 operators\,\cite{Degrande_2014}. 
Hence nTGCs can provide a unique window for probing new physics beyond the SM\,
\cite{NewEFTnTGC,Ellis_2020,Ellis_2023_lly,Ellis_2023_nny,Jahedi:2022duc,Jahedi:2023myu}.\  
The most direct experimental probes of nTGCs are via measurements of the corresponding form factors. A consistent formulation of nTGC form factors has recently been proposed, which matches precisely 
the nTGC form factors with the gauge-invariant dimension-8 effective operators of the SMEFT\,\cite{Ellis_2023_lly,Ellis_2023_nny}.\ 
This imposes nontrivial relations among the nTGC form factors and 
gives correct predictions for the contributions of the nTGC form factors to high-energy scattering amplitudes\,\cite{Ellis_2023_lly,Ellis_2023_nny}.\  
These theoretical papers investigated probes of the nTGCs at both the electron-positron and hadron colliders.

\begin{figure}[!ht]
    \centering
    \includegraphics[width=.3\columnwidth,height=.22\columnwidth]{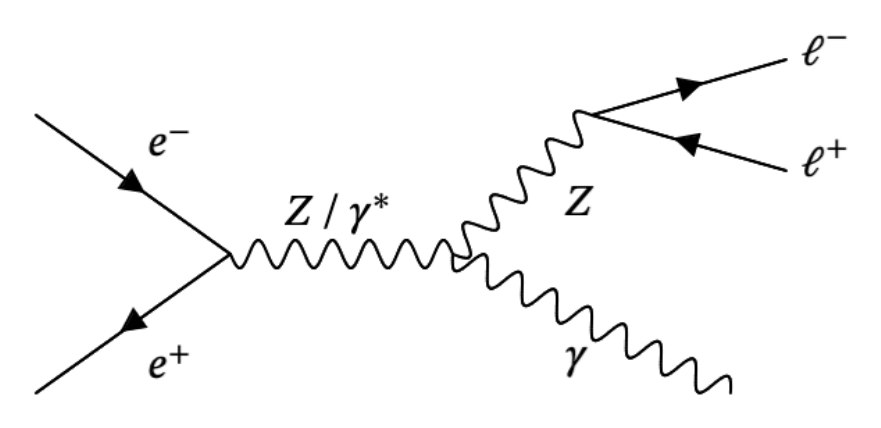}
    \includegraphics[width=.3\columnwidth,height=.22\columnwidth]{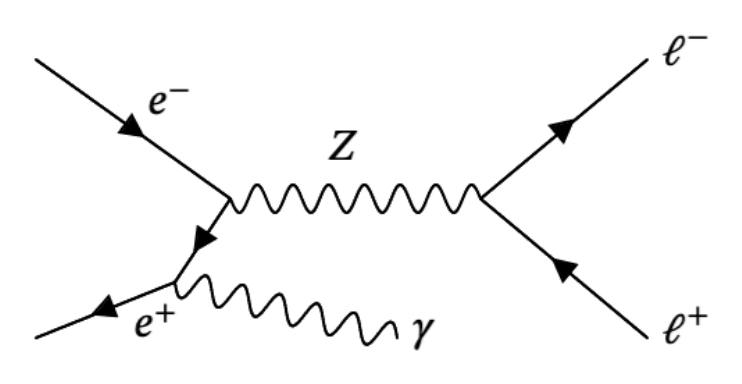}  
    \includegraphics[width=.3\columnwidth,height=.22\columnwidth]{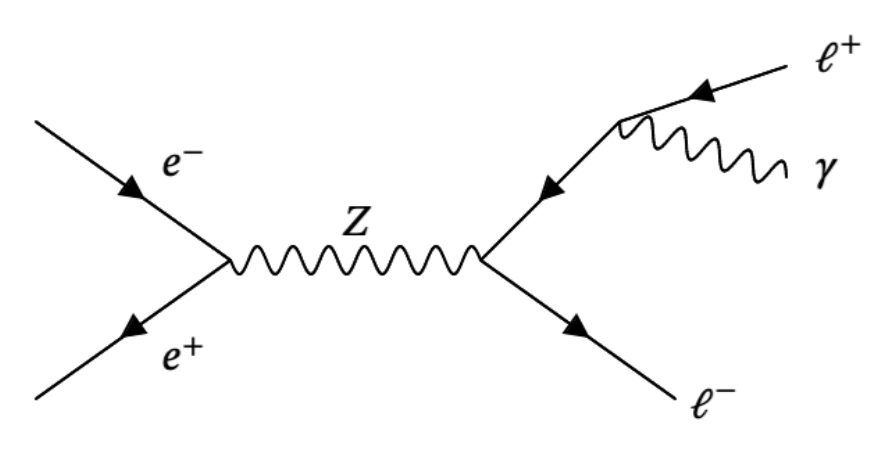}
\caption{\small Feynman diagrams that contribute to the reaction $e^+e^-\!\rightarrow Z\gamma\,$.\ The first diagram is the signal process containing the nTGC vertex $Z^* Z\gamma$ or $\gamma^* Z\gamma$; the second and third diagrams 
show the SM background contributions with initial-state-radiation photon or 
final-state-radiation photon.}
\label{fig:FeynmanDiagrams_Zy}
\label{fig:1}
\end{figure}

In this work, we study experimental
probes of the dimension-8 nTGC operators via measurements of their corresponding nTGC form factors in the reaction $e^+e^-\!\rightarrow\! Z\gamma$ process with $Z\!\rightarrow\! \ell^+\ell^-$ 
$(\ell =e,\mu)$ decays, as shown in Fig.\,\ref{fig:FeynmanDiagrams_Zy}.\ 
For this purpose we perform detector-level simulation and
analysis of nTGCs at the Circular Electron Positron Collider
(CEPC) with energy $\sqrt{s} = 240$~GeV and an integrated luminosity of 20~ab$^{-1}$, using a model-independent approach that could also be adopted for other experiments.\

This work is organized as follows.\ 
In Section\,\ref{sec:simuandana}, we first describe the theoretical framework for the nTGCs, 
which includes the SMEFT formulation of the dimension-8 nTGC operators
and the corresponding nTGC form factors.\ 
Then, we present a detector-level simulation 
and analysis for the dimension-8 nTGC effective operators   
and the nTGC form factors via the reaction
$e^+e^-\!\!\to\! Z\gamma$, using the CEPC detector as a benchmark.\ 
In Section\,\ref{sec:3}, we analyze the uncertainties for both the signals and
backgrounds.\ After this, we present our results in Section\,\ref{sec:4}
for the sensitivities on probing the nTGCs at the CEPC.\ 
Finally, we conclude in Section\,\ref{sec:5}.  

%% file: Combined_Chapter2.tex

In this Section we first present the theoretical framework for the nTGCs, 
including both the SMEFT formulation with dimension-8 nTGC operators
and the corresponding nTGC form factors.\ 
Then, we systematically perform a detector-level simulation 
and analysis for the dimension-8 nTGC effective operators   
and the corresponding nTGC form factors 
via the reaction
$e^+e^-\!\!\to\! Z\gamma$\,, using the CEPC as a benchmark.

\subsection{Theoretical Framework for the nTGCs}

The dimension-8 SMEFT effective Lagrangian takes the following form:
\begin{equation}
\mathcal{L}_{\rm{SMEFT}}^{} \,=\, 
\sum_{j}\frac{c_{j}}{\,\Lambda^{4}\,}\mathcal{O}_{j}^{} 
\,=\, 
\sum_{j}\frac{\,\text{sign}({c}_{j}^{})\,}{\,\Lambda^{4}_j\,}\mathcal{O}_{j}^{}\,,
\label{eq:Lagrangian_eft}
\end{equation}
where the $\{{c}_{j}^{}\}$ are dimensionless coefficients that may be $\mathcal{O}(1)$ that can have either sign.\
The effective cutoff $\Lambda$ for the new physics scale is connected to
$\Lambda_j^{}$ via $\,\Lambda_{j}^{} \!\equiv\! {\Lambda}/|{c}_{j}|^{1/4}\,$. 

In the present analysis we consider the following set of CP-conserving  
dimension-8 nTGC operators 
($\mathcal{O}_{G+}, \,\mathcal{O}_{G-},\, \mathcal{O}_{\tilde{B}W},\, 
  \mathcal{O}_{\widetilde{B W}}$) 
\cite{Ellis_2020,Ellis_2023_lly,Ellis_2023_nny}: 
\begin{subequations}
\begin{align}
g\mathcal{O}_{G+} &= \tilde{B}_{\mu\nu}W^{\alpha\mu\rho}(D_{\rho}D_{\lambda}W^{\alpha\nu\lambda}+D^{\nu}D^{\lambda}W^{\alpha}_{\lambda\rho}) \,,
\label{eq:OGplus_expression}
\\
g\mathcal{O}_{G-} &=\widetilde{B}_{\mu \nu} W^{a \mu \rho}
(D_\rho D_\lambda W^{a \nu \lambda}-D^\nu D^\lambda W_{\lambda \rho}^a) \,,
\label{eq:OGminus_expression}
\\
\mathcal{O}_{\widetilde{B} W} &= \mathrm{i} H^{\dagger} \widetilde{B}_{\mu \nu} W^{\mu \rho}\left\{D_\rho, D^\nu\right\}\! H+\rm{h.c.}  \,,
\label{eq:OBtW_expression}
\\
\mathcal{O}_{\widetilde{B W}}^{} &=
\mathrm{i} H^{\dagger} \big(  D_{\!\sigma}^{}{\widetilde W}^a_{\!\mu\nu}W^{a\mu\sigma}\!\!+\! D_{\!\sigma}^{}{\widetilde B}_{\mu\nu}B^{\mu\sigma}\big)D^\nu\!H\!+\!\text{h.c.}
\label{eq:OBW_expression}
\end{align}
\end{subequations}
The nTGC vertex $Z\gamma V^*$ ($V=Z,\gamma$) can be expressed in terms of
nTGC form factors $(h_3^V,\,h_4^V)$ as follows\,\cite{Ellis_2023_lly,Ellis_2023_nny}:
\begin{equation}
\Gamma^{\alpha\beta\mu(8)}_{Z\gamma V*}(q_1^{}, q_2^{}, q_3^{}) 
= \frac{\,e(q_3^2-\!M_v^2)\,}{M_{Z}^2}
\!\left[\!\left(\!h_3^V + \frac{h_{4}^{V}}{\,2M_{Z}^{2}\,}q_3^2\right)\!
q_{2\nu}^{}\epsilon^{\alpha\beta\mu\nu} 
\!+ \frac{h_4^V}{\,M_{Z}^{2}\,} q_2^{\alpha}q_{3\nu}q_{2\sigma}
\epsilon^{\beta\mu\nu\alpha}\right] \!.
    \label{eq:ZyVertex_Expression_02}
\end{equation}
By matching this nTGC form factor formulation with the corresponding 
gauge-invariant dimension-8 nTGC operators, a nontrivial 
form factor relationship can be derived,
$\,h_{4}^{Z} \!=\!\frac{c_W^{}}{s_W^{}}h_{4}^{\gamma}$ 
\cite{Ellis_2023_lly,Ellis_2023_nny}, and henceforth we will denote 
$h_4^{Z}\!\equiv\!h_{4}^{}$ for simplicity.\ 
Thus there are three independent form-factor parameters 
($h_{4}^{}, h_{3}^{Z}, h_{3}^{\gamma}$)
\cite{Ellis_2023_lly,Ellis_2023_nny}, 
which can be determined by matching the gauge-invariant dimension-8
nTGC operators ($\mathcal{O}_{G+}, \mathcal{O}_{G-}, \mathcal{O}_{\tilde{B}W}, \mathcal{O}_{\widetilde{B W}}$) in the broken phase of the electroweak gauge group 
SU(2)$_{\!L}^{}\otimes$U(1)$_Y^{}$.\ 
The form factors ($h_{4}, h_{3}^{Z}, h_{3}^{\gamma}$) are connected as follows to the cutoff scales ($\Lambda_{G+}, \Lambda_{G-}, \Lambda_{\tilde{B}W}, \Lambda_{\widetilde{B W}}$)
of the corresponding dimension-8 nTGC operators\,\cite{Ellis_2023_lly,Ellis_2023_nny}:
\begin{subequations}
\label{eq:h_conversion}
\begin{align}
h_{4} &=- \frac{1}{[\Lambda^{4}_{G+}]}\frac{v^2M_{Z}^2}{s_{W}c_{W}} \,,
        \label{eq:h4_conversion}\\
h_{3}^{Z} &=\frac{1}{[\Lambda^4_{\tilde{B}W}]}\frac{v^2M_{Z}^2}{2s_{W}c_{W}} \,,
        \label{eq:h3Z_conversion}\\
h_{3}^{\gamma} &= -\frac{1}{[\Lambda^4_{G-}]}\frac{v^2M_{Z}^2}{2c_W^2}  
        =-\frac{1}{[\Lambda^4_{\widetilde{B W}}]}\frac{\,v^2M_Z^2\,}{\,s_W^{}c_W^{}\,} \,, 
\label{eq:h3y_conversion}
\end{align}
\end{subequations}
where we denote $[\Lambda_j^4]=\text{sign}(c_j)\Lambda_j^4\,$
and $[\Lambda_j^{-4}]=\text{sign}(c_j)\Lambda_j^{-4}\,$.\ 

In the following, we perform a systematic detector-level simulation and analysis 
of sensitivities to the nTGC dimension-8 effective operators 
via measurements of the nTGC form factors in the reaction
$e^+e^-\!\to\! Z\gamma$ at the CEPC. 

\subsection{CEPC Detector}

The Circular Electron Positron Collider (CEPC)\,\cite{CEPCcdr, CEPCStudyGroup:2023quu}  is an international research facility proposed in China that is designed to meet the requirements of various physics studies, especially
precision measurements.\ 
CEPC has well-defined momentum and energy, as well as a clean experimental environment in comparison with hadron colliders.\ 
The energy resolution for the electromagnetic calorimeter is at $16\% \sqrt{\text{E/GeV}} \oplus 1\%$, and for the hadronic calorimeter and muon detector is at $60\% \sqrt{\text{E/GeV}} \oplus 1\%$. The angular resolution is set to be less than 0.1 $\text{mrad}$\cite{CEPCcdr, Zhang:2024bld}.
Thus it is possible to reconstruct angular variables in a more accurate way.\ Hence, CEPC is an ideal facility for probing  
new physics beyond the SM.

\subsection{Simulation}
For the purpose of this analysis, signal events are generated using \textsc{MadGraph5}\_aMc$@$NLO~\cite{IntroductionMG5} and \textsc{Pythia8}~\cite{IntroductionPY8}, using the nTGC formulation described in Section~\ref{sec:intro}. This nTGC formulation is implemented and imported to \textsc{MadGraph5}\_aMc$@$NLO using FeynRules for nTGC event production at the matrix element level at leading order. \textsc{Pythia8} is used for parton showering, fragmentation and describing the underlying events.

\begin{figure}[!ht]
    \centering
    \includegraphics[width=.42\columnwidth]{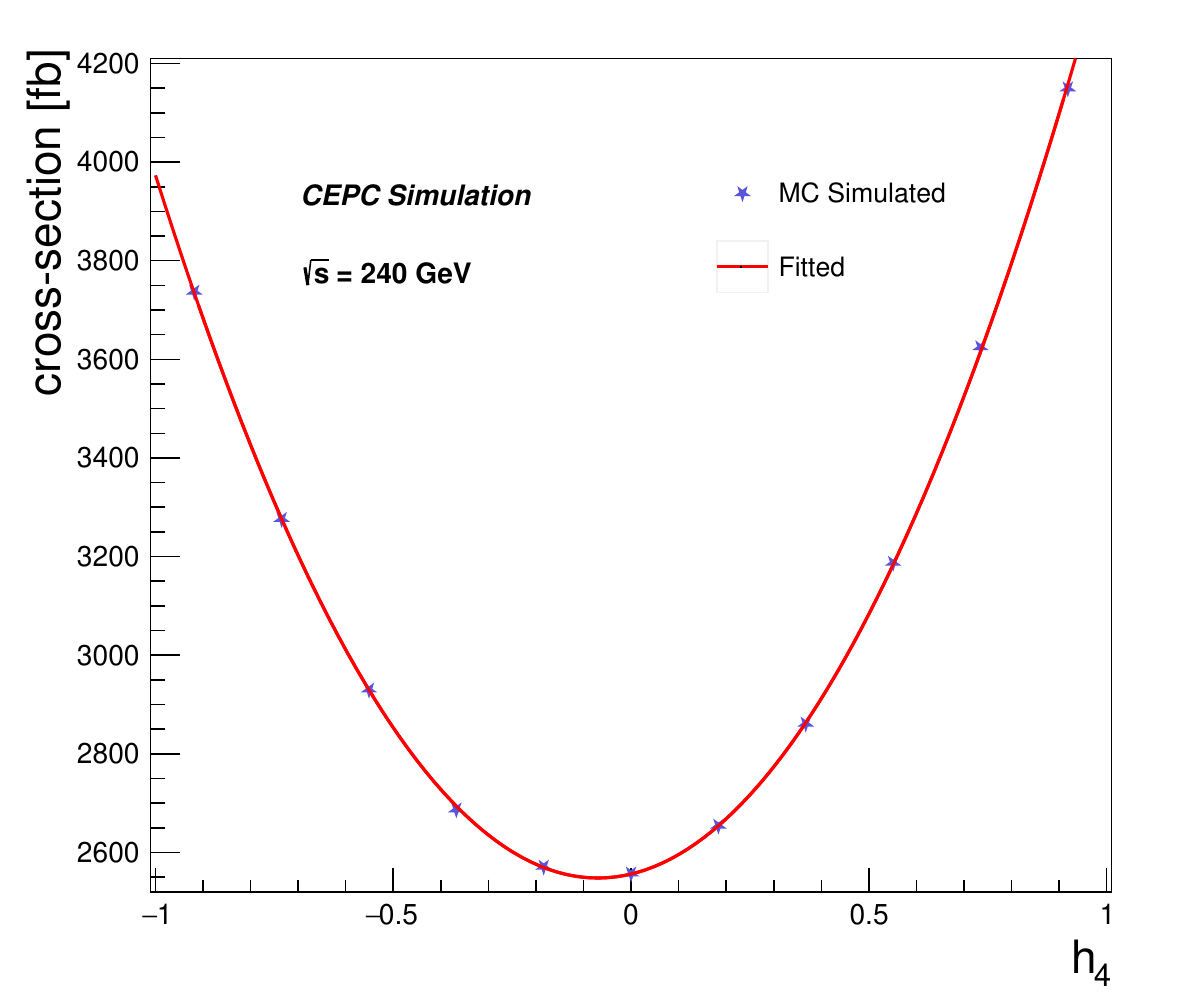}
    \includegraphics[width=.42\columnwidth]{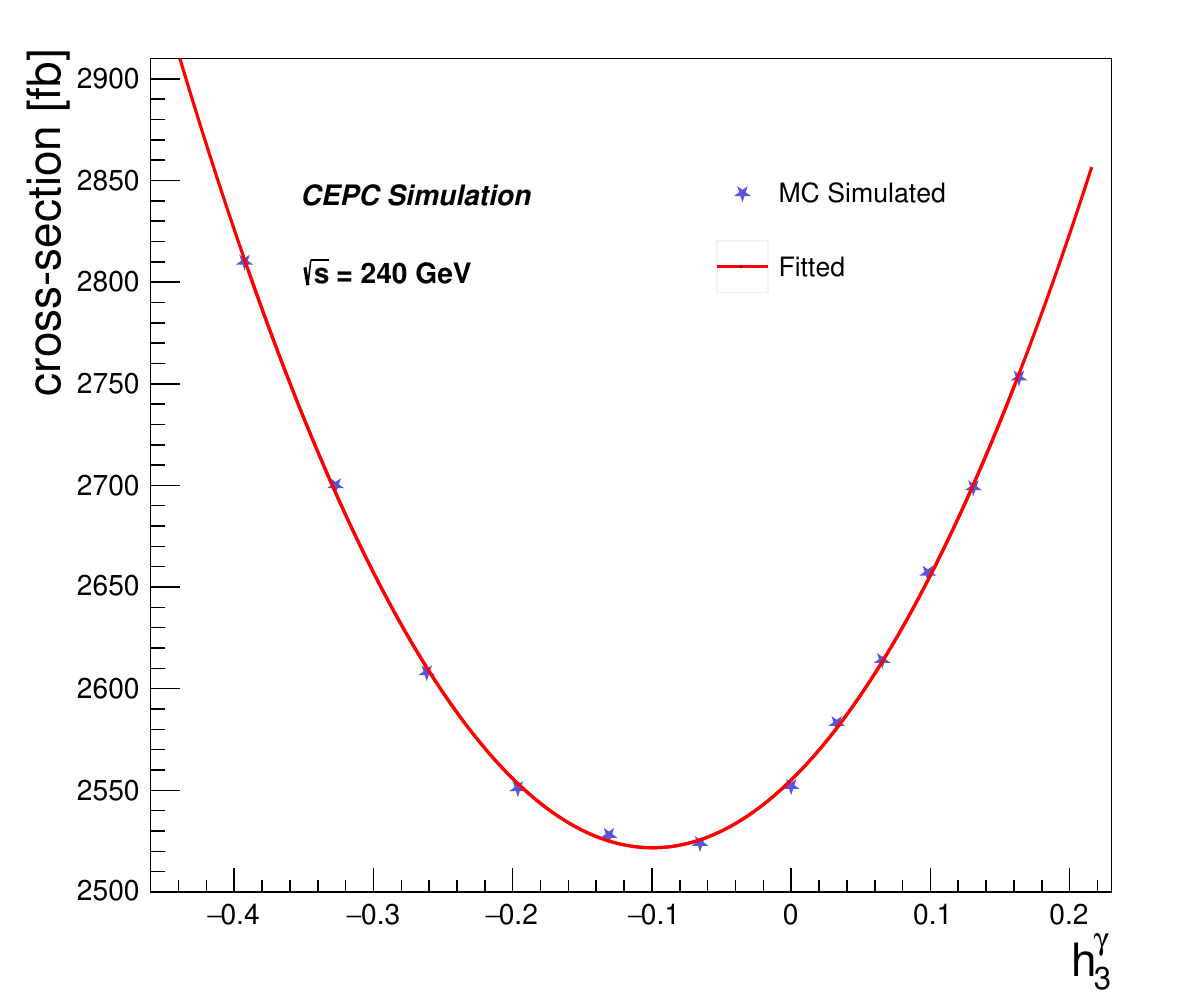}
    \includegraphics[width=.42\columnwidth]{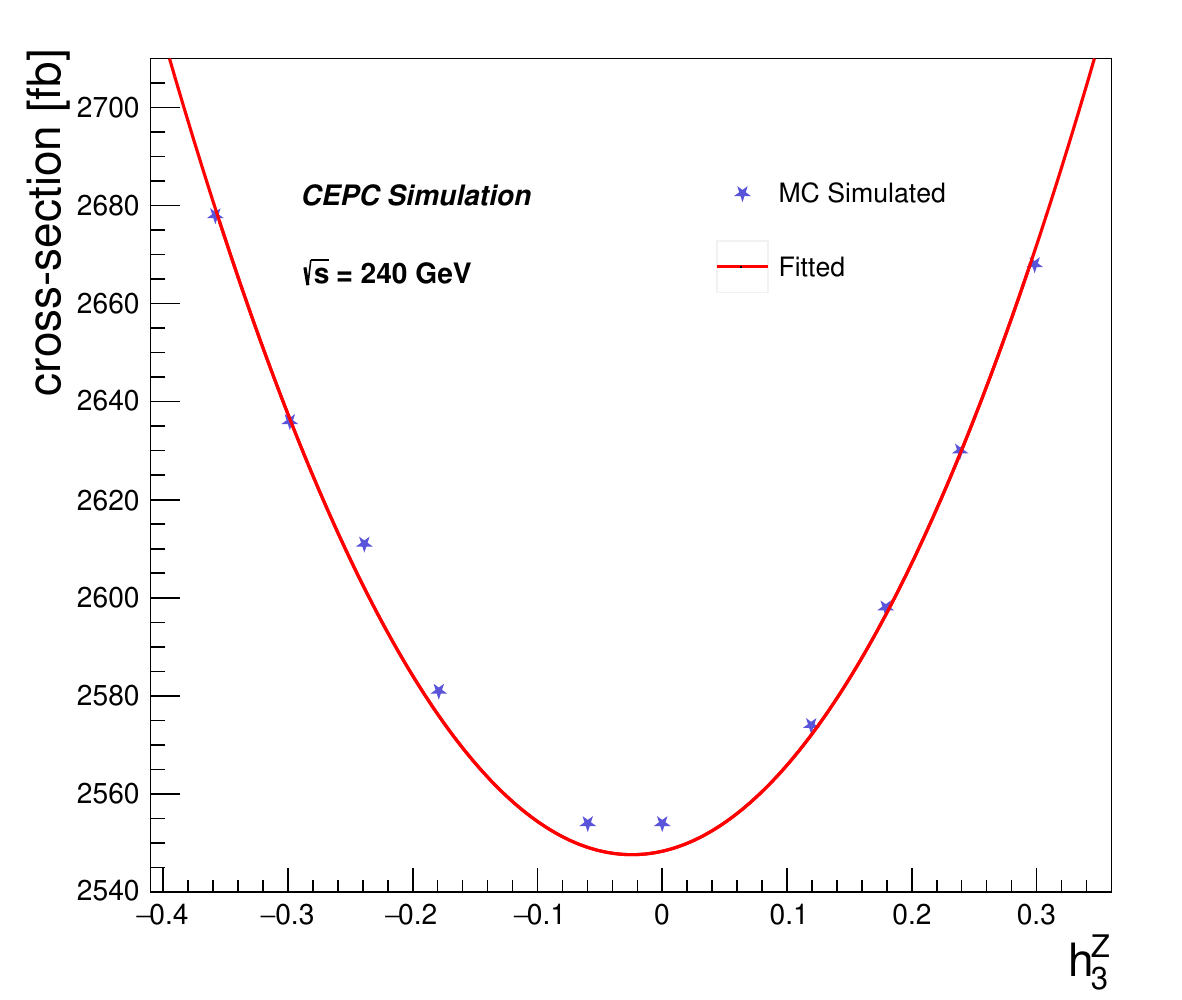}
\vspace*{-3.5mm} 
\caption{\small {%
Cross sections as functions of the form factors $h_{4}, h_{3}^{\gamma}$, and $h_{3}^{Z}$.
The interrelation between cross sections and form factors is evaluated by varying the parameter values, which agree well with the polynomial form of Eq.~\ref{eq:onedimensional_xs}. The solid blue stars represent the outputs from \textsc{MadGraph}\_aMc$@$NLO simulations, while the red solid curves indicate the fits extracted through parametrizations corresponding to Eq.~\ref{eq:onedimensional_xs}.}} 
\label{fig:onedimensional_xsplots}
\end{figure}

We illustrate the contributions of the
three nTGC form factors $(h_4^{}\,\hs h_3^Z,\hs h_3^{\gamma})$ 
with the benchmark choices shown in the second row of Table\,\ref{tab:InitialValues}, whose corresponding  
cross sections are shown in its third row.\   
The dependence of the $Z\gamma$ cross section 
on the nTGC form factors $h_j^{}$ (and the corresponding cutoff
scale $\Lambda_j^{}$) can be expressed as follows:
%
\begin{align}
\sigma^{}_{Z\gamma}
= \sigma_0^{} + 
\bar{\sigma}_{1}^{}h_j^{} + 
\bar{\sigma}_{2}^{}h_j^2 
= \sigma_0^{} + 
\tilde{\sigma}_{1}^{}[\Lambda^{-4}_j] + 
\tilde{\sigma}_{2}^{}\Lambda^{-8}_j \,,
\label{eq:onedimensional_xs}  
\end{align}
%
where $\sigma_0^{}$ is the SM contribution, 
$\bar{\sigma}_{1}^{}$ or $\tilde{\sigma}_{1}^{}$
arises from the interference term between the nTGC and SM
contributions, and $\bar{\sigma}_{2}^{}$ or
$\tilde{\sigma}_{2}^{}$ corresponds to the squared 
nTGC contributions.
In the above we use the notation
$[\Lambda_j^{-4}]=\text{sign}(c_j)\Lambda_j^{-4}\,$
as defined below Eq.\eqref{eq:h_conversion}. 
Fig.~\ref{fig:onedimensional_xsplots} presents results obtained by scanning various form factors ($h_{4}$, $h_{3}^{Z}$, $h_{3}^{\gamma}$).
The fitted curves in these plots indicate that the cross sections agree well with Eq.~\ref{eq:onedimensional_xs}, 
confirming the dependence of the $Z\gamma$ cross section with the nTGC form factors.

\begin{table}[!ht]
    \centering
    \begin{tabular}{c| c c c c c c c}
    \toprule
    Form Factors                    & SM     & $h_{4}$  & $h_{3}^{\gamma}$ & $h_{3}^{Z}$ & $(h_{4},\,h_{3}^{\gamma})$ & $(h_{4},\,h_{3}^{\gamma})$ & $(h_{3}^{\gamma},\,h_{3}^{Z})$ \\
    \midrule
    $h_{i}^{V}$                     & -      & 0.28     & 0.16             & 0.36        & (0.83, 0.49)  & (0.83, 1.07) & (0.49, 1.07) \\
    $\sigma_{Z\gamma}^{}$ (fb)      & 2551.7 & 2616     & 2752             & 2712        & 3732          & 3613         & 5120         \\
    \bottomrule
    \end{tabular}
\caption{\small Benchmark values for the form factors
$(h_4^{}\,\hs h_3^Z,\hs h_3^{\gamma})$ (second row) 
and the corresponding cross sections for $Z\gamma$ production (third row).}
\label{tab:InitialValues}
\label{tab:1}
\end{table}

In Table\,\ref{tab:1},
each nTGC benchmark consists of three contributions: 
the SM term, the interference term between the SM and nTGC, 
and the squared nTGC term.\ 
We achieve accurate sample production by decomposing the 
$Z\gamma$ cross section into these three terms and generating each term independently.\ 

The package 
\texttt{WHIZARD}\,\cite{IntroductionWhizard} is used to simulate background events.\ All background samples considered in this analysis can be divided into three categories:\ the 2-fermion background (which is dominant),  the 4-fermion backgrounds,
and the resonant Higgs backgrounds.\ Detailed information on the background processes is given in 
Table\,\ref{tab:mcsample_backgrounds} of the Appendix.\ 

\vspace*{1mm} 

The simulation of the detector response is handled by MokkaPlus~\cite{IntroductionMokka}, a \textsc{GEANT4}~\cite{IntroductionG4}-based framework.\ 
We perform the full detector simulation for the signal process, whereas the background processes are simulated using Delphes\,\cite{IntroductionDelphes}.

\subsection{Analysis Strategy}

The CEPC detector adopts the Particle Flow Algorithm (PFA)~\cite{IntroductionPFA_Manqi} for event reconstruction, using the dedicated toolkit Arbor~\cite{IntroductionArbor}, which collects tracks and hits from the calorimeter and composes the Particle Flow Objects (PFOs) with its clustering and matching modules. The CEPC detector acts like a ``camera" that tracks every particle collision. It is not possible to observe all the particles directly in the collisions because some of them decay promptly or do not interact with the detector. However, if they decay to stable particles or interact with the apparatus, they  leave signals in the subdetectors. These signals are used to reconstruct the decay products or to infer their presence as physics objects. These objects can be photons, electrons, muons, jets, missing energy, etc. 

In this analysis, photons are identified in Arbor using shower shape variables obtained from the high granularity calorimeter without any matched tracks. Leptons ($e^{\pm}, \mu^{\pm}$) are identified by a track-matched particle. A likelihood-based algorithm, LICH \cite{LeptonID}, is implemented in Arbor to separate electrons, muons, and hadrons. The overall lepton identification efficiencies~\cite{LeptonID} for electrons and muons are 99.7\% and 99.9\% respectively, where mis-identification rates are lower than 0.07\%. To reconstruct fully electrons and muons, and to make sure no ambiguity exists, a lepton isolation criterion~\cite{CEPCZH_BaiYu} is also applied by requiring $E_{\text{cone}}^{2} < 4E_{\ell} + 12.2$, where $E_{\text{cone}}$ is the energy within a cone with $\cos\theta_{\text{cone}} < 0.98$ around the lepton and $E_{\ell}$ is the energy of the lepton. Here $E_{\ell}$ and $E_{\text{cone}}$ are measured in GeV. The polar angle between two selected leptons systems is required to be within the range $|\cos\theta_{\mu^+\mu^-}| < 0.81$ and $|\cos\theta_{e^+e^-}| < 0.71$ so as to ensure that the selected leptons are isolated. Jets are also reconstructed by Arbor, after removing isolated leptons and photons so as to avoid mis-reconstruction due to lepton or photon constituents. A list of object definitions is shown in Table~\ref{tab:ObjReco_table}.

\begin{figure}[!ht]
\centering
\includegraphics[width=.49\columnwidth]{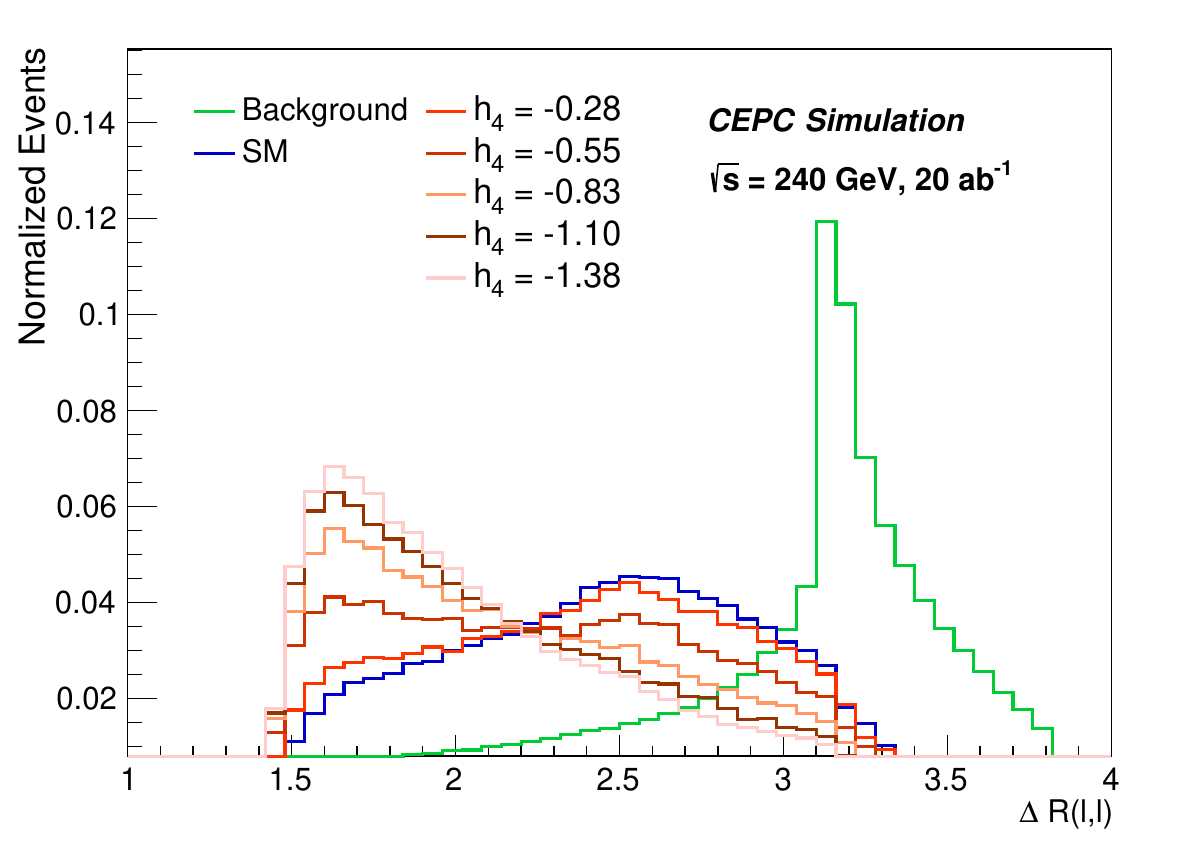}
\includegraphics[width=.49\columnwidth]{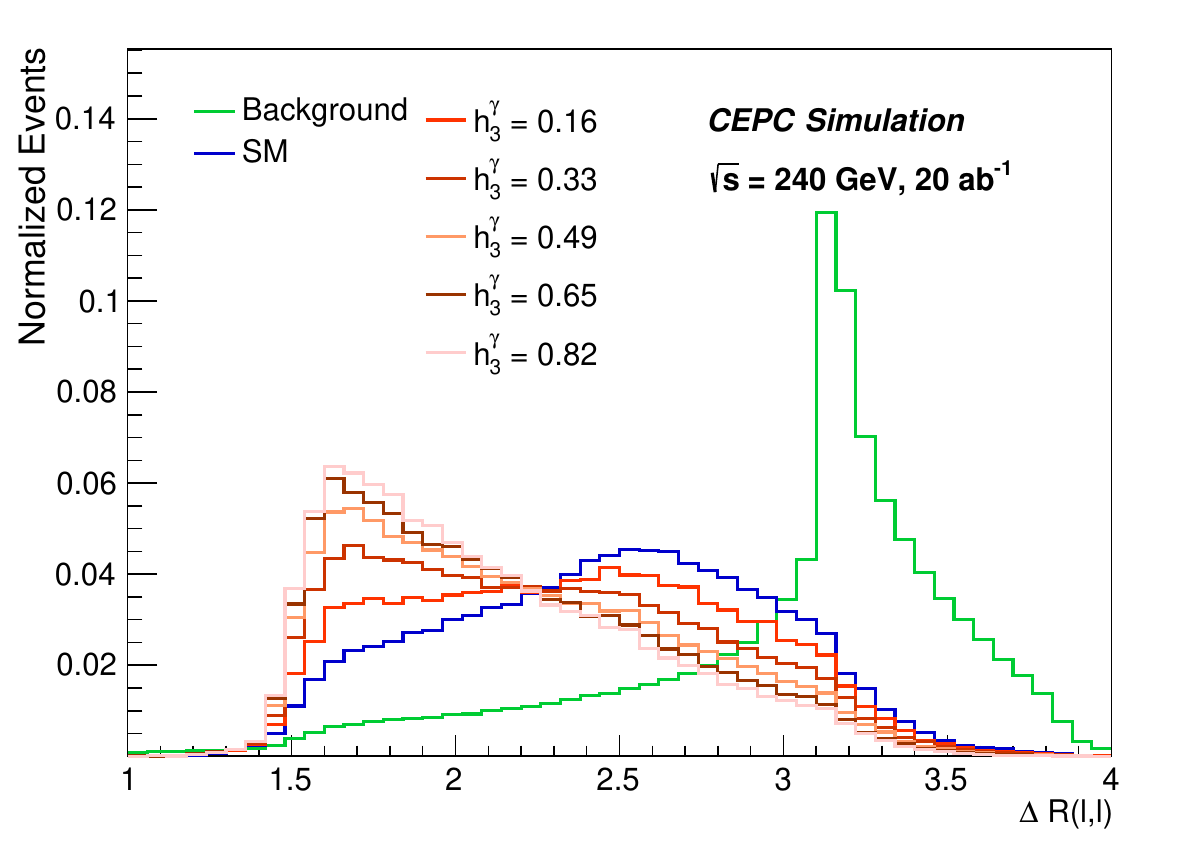}    \\
\includegraphics[width=.49\columnwidth]{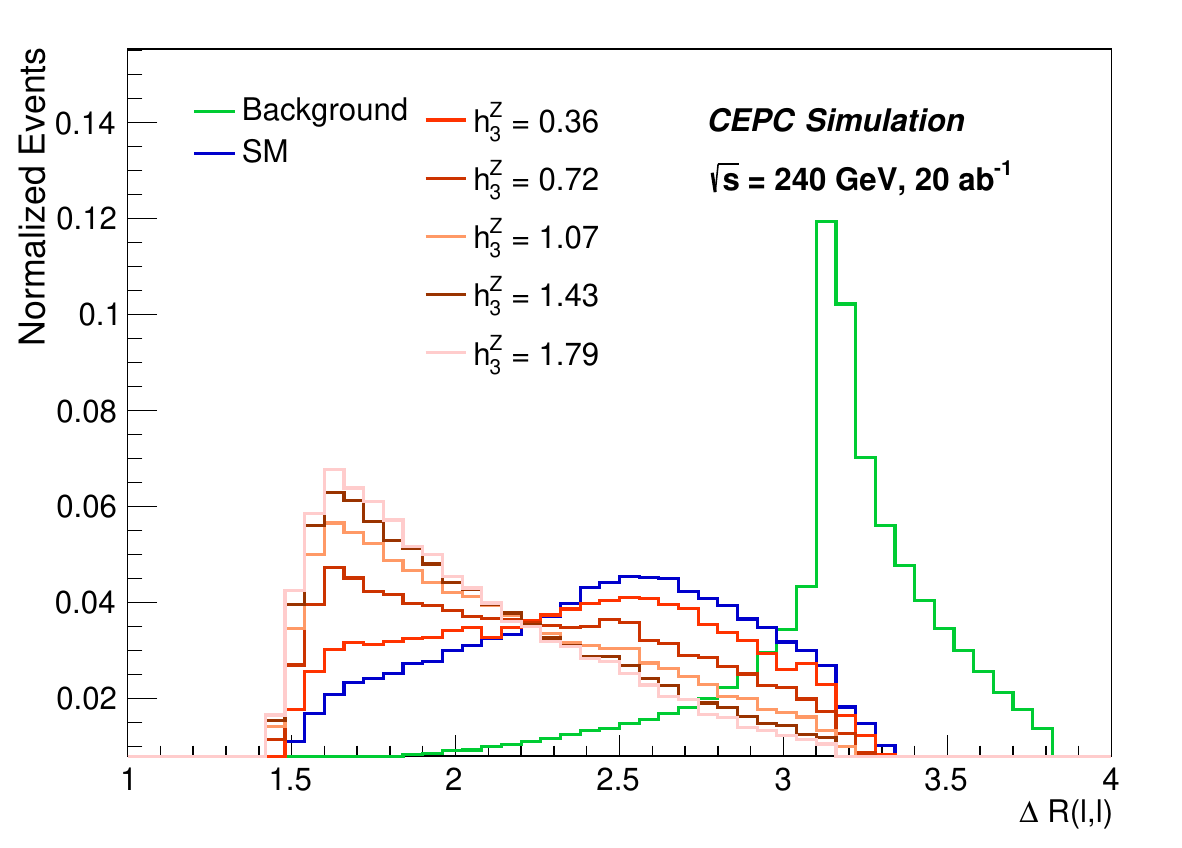}   
\\
\vspace*{-3.5mm} 
\caption{\small Normalized distributions of the separation $\Delta R(\ell, \ell)$, including the effects of different dimension-8 operators.
             The performance of $\Delta R(\ell, \ell)$ for leptons originating from different sources exhibits significant variations.
             The distinctions in $\Delta R(\ell, \ell)$ between different processes play pivotal roles in enhancing signal detection and minimizing background contributions.}
\label{fig:kinematic_distribution}
\vspace*{5mm}
\end{figure}

\begin{table}[!ht]
    \centering
    \begin{tabular}{c c}
        \toprule
        Objects    &   Requirements    \\
        \midrule
        Electrons   &   $p_{T} >$ 15 GeV, $|\cos\theta| < $0.969 \\
                    &   $E^2_{\text{cone}} < 4E_{\ell} + 12.2$  \\
                    &   $|\cos\theta_{e^+e^-}| <$ 0.71   \\
                    &   (Overlap removal) $\Delta R(e,j) >$ 0.4, $\Delta R(e,\mu) >$ 0.4\\
        \midrule
        Muons       &   $p_{T} >$ 15 GeV, $|\cos\theta| < $0.969\\
                    &   $E^2_{\text{cone}} < 4E_{\ell} + 12.2$  \\
                    &   $|\cos\theta_{\mu^+\mu^-}| <$ 0.81   \\
        \midrule
        Photons     &   $p_{T} >$ 30 GeV, $|\cos\theta| < $0.969\\
                    &   (Overlap removal) $\Delta R(\gamma, e) >$ 0.4, $\Delta R(\gamma, \mu) >$ 0.4    \\
        \midrule
        Jets        &   $p_{T} >$ 25 GeV, $|\cos\theta| < $0.969\\
                    &   (Overlap removal) $\Delta R(j,\gamma) >$ 0.4, $\Delta R(j, e) >$ 0.4    \\
        \bottomrule
\end{tabular}
\caption{\small Summary of selection cuts on leptons, photons, and jets.\ These basic cuts\,\cite{CEPCZH_BaiYu} are independent of generator implementation and are needed for selecting stable particles as well as the analysis of complex event topologies.}
\label{tab:ObjReco_table}
\label{tab:2}
\end{table}

This analysis is based on events with just one photon and a pair of leptons of the same flavor and opposite signs (electron and muon). Events with more than one photon or a pair of charged leptons are vetoed. 
The event selections summarised in Table~\ref{tab:EventSel_table} are applied to improve the signal significance. 

The event selections are optimised according to the requirements of the formulation in~\cite{NewEFTnTGC}. We first request that no selected jets be left in the signal events, so as to remove higher-order corrections appearing at Next-to-Leading Order (NLO) and beyond as much as possible, and to ensure that cross section enhancement comes from nTGC, not higher-order SM corrections or other SM jet backgrounds. This is an effective cut to remove other SM backgrounds and to improve sensitivity. In this scenario, we also require that two leptons must come from the same Z boson by requiring the invariant mass difference between the di-lepton system and the on-shell Z boson mass be smaller than 10 GeV. Events with final-state radiation photons (FSR) are suppressed by requiring that the sum of the invariant mass of the leptons and the invariant mass of leptons and photon is greater than twice the Z mass ($|m_{\ell\ell} + m_{ll\gamma}| > $182 GeV). We also apply the cut $\Delta R(\ell,\ell) < 3$ to suppress background contributions as shown in Fig.~\ref{fig:kinematic_distribution}. All the selections listed in Table~\ref{tab:EventSel_table} are required so as to make the correct transformation between the SMEFT and the Effective Vertex Theory formulated in~\cite{NewEFTnTGC}.

\begin{table}[!ht]
    \centering
    \begin{tabular}{c c}
        \toprule
        Variables   &   Cut \\
        \midrule
        $N_{\rm{lep}}^{}$   &   2 signal OSSF leptons with leading lepton $p_{T}^{\rm{lep}} > 30$ GeV  \\
        $N_{\rm{pho}}^{}$   &   $\ge$ 1 signal photon with $p_{T}^{\gamma} > 35$ GeV  \\
        $N_{\rm{jet}}^{}$   &   0   \\
        $\Delta R(\ell,\ell)$ &   $< 3$   \\
        $m_{\ell\ell}$    &   $|m_{\ell\ell} - m_{Z}| < 10$ GeV \\
        $m_{\ell\ell} + m_{\ell\ell\gamma}$ &   $> 182$ GeV \\
        \bottomrule
    \end{tabular}
    \caption{\small Summary of event selection cuts used in this analysis.}
\label{tab:EventSel_table}
\label{tab:3}
\end{table}

In this measurement the nTGC form factors are constrained 
by measurements of $e^+e^-\!\to Z\,(\rightarrow \ell^+\ell^-)\gamma$
where $\ell = e, \mu$.\ 
An event selection strategy is proposed based on
the new form factor formulation and summarised in Tables\,\ref{tab:CutFlow_table_01} and \ref{tab:CutFlow_table_02}, which display the signal cut-flow results including 
contributions of the SM, the interference term, and the quadratic term. 

\begin{table}[!ht]
    \centering
    \begin{tabular}{c  c  c  c  c  c}
    \toprule
        Variables                           & SM Backgrounds & SM $Z\!\gamma$ & $h_{4}$  & $h_{3}^{\gamma}$ & $h_{3}^{Z}$     \\ 
 \midrule
        $N_{\rm{pho}}^{} \geqq 1$           &  11712  & 1572   & 1629   & 1747   & 1710   \\
        $N_{\rm{lep}}^{} = 2$               &  1152   & 587    & 624    & 696    & 675    \\
        $N_{\rm{jet}}^{} = 0$               &  811    & 587    & 624    & 696    & 675    \\
        $\Delta R(\ell,\ell) \!<\! 3$         &  698   & 548    & 585   & 656   & 634   \\
        $|m_{\ell\ell}^{} \!-\! m_{Z}| \!<\! 10$\,GeV &  303   & 192    & 226   
        & 288   & 271  \\
$(m_{\ell\ell}^{} \!+\! m_{\ell\ell\gamma}^{}) \!>\! 182$\,GeV  & 300 & 192 & 226 & 288   & 271   \\
    \bottomrule
    \end{tabular}
\caption{\small Cut-flow table for the nTGC form factors, enumerating the cross sections (in fb) after applying sequential selections and using the indicated event-topology requirements.\ 
The initial cross sections for each nTGC form factor are shown 
in Table\,\ref{tab:InitialValues}.\ 
The implementation of these selections mitigates SM background contributions efficiently, whereas it preserves signal events. }
\label{tab:CutFlow_table_01}
\end{table}

\begin{table}[!ht]
    \centering
    \begin{tabular}{c  c  c  c  c  c}
    \toprule
        Variables                   & SM Backgrounds & SM $Z\!\gamma$ & $(h_{4}, h_{3}^{\gamma})$  & $(h_{4}, h_{3}^{Z})$ & $(h_{3}^{\gamma}, h_{3}^{Z})$     \\ 
 \midrule
        $N_{\rm{pho}}^{} \geqq 1$               &  11712  & 1572   & 2614   & 2506   & 3811   \\
        $N_{\rm{lep}}^{} = 2$                   &  1152   & 587    & 1225   & 1178   & 1999   \\
        $N_{\rm{jet}}^{} = 0$                   &  811    & 587    & 1224   & 1176   & 1996   \\
        $\Delta R(\ell,\ell) \!<\! 3$           &  698   & 548    & 1179   & 1126   & 1929   \\
        $|m_{\ell\ell}^{} \!-\! m_{Z}| \!<\! 10$\,GeV &  303   & 192    & 751    & 717    & 1441   \\
        $(m_{\ell\ell}^{} \!+\! m_{\ell\ell\gamma}^{}) \!>\! 182$\,GeV  & 300 & 192 & 751  & 717   & 1441   \\
    \bottomrule
    \end{tabular}
\caption{\small Cut-flow table for pairs of nTGC form factors, enumerating the cross sections (in fb) after applying sequential selections and using the indicated event-topology requirements.\ 
The initial cross sections for the pairs of nTGC form factors are shown in Table\,\ref{tab:InitialValues}.\ 
The implementation of these selections efficiently mitigates SM background contributions, whereas preserving signal events.}
\label{tab:CutFlow_table_02}
\end{table}

%% file: Systematics.tex

We have considered several sources of systematic uncertainties, which can be grouped into two types: theoretical and experimental uncertainties. Both systematic uncertainties have been assigned to the expected signal yields and then propagated to the SMEFT fits. 
 
\subsection{Signal Uncertainties}

Unlike hadron colliders, only a few theoretical uncertainties influence the final measurement in lepton colliders such as CEPC. There is no impact from Parton Distribution Functions or $\alpha_{s}$, and little dependence on higher-order QCD corrections. For completeness, a 0.5\% theoretical uncertainty~\cite{CEPCyy_FY} is assumed for the signal yields.

The experimental systematic uncertainties include those in the integrated luminosity, detector acceptance, trigger efficiency, object reconstruction and identification efficiency, object energy scale, and resolution. Luminosity in the CEPC detector is monitored by the LumiCal using the high-statistics BhaBha process, and a relative accuracy of 0.1\% is expected to be achieved~\cite{CEPCyy_FY}. A well-described detector geometry is used in the simulation  to provide a precise model of the detector acceptance and response. These uncertainties should be negligible in our analysis. The photon identification, reconstruction, and energy calibration rely on dedicated algorithms and real data. All these photon-related uncertainties are detailed and studied in the CEPC CDR~\cite{CEPCcdr} and controlled at the sub-percent level. We assume conservatively a 1\% uncertainty in the the photon efficiency and 0.05\% uncertainties \cite{CEPCyy_FY} in the photon energy scale (PES) and resolution (PER). The lepton uncertainties are estimated by varying the Z boson mass selection by $\pm 1$~GeV. The differences between the varied and nominal signal yields will be considered lepton uncertainties, which are strongly related to the lepton selection criteria. 

\subsection{Background Uncertainties}

The background yields are floated to consider background mis-modelling effects and uncertainties in cross section calculations.\ Fixed parameters are used to estimate uncertainties from different background processes.\ 
The event yields of the dominant 2-fermion background process are varied by $\pm 5$\%, and the yields from other background processes (4 fermions and Higgs production) are varied by $\pm 100$\%. These estimates are based on the recipe described in~\cite{CEPCZH_BaiYu}.

\begin{table}[!h]
    \centering
    \begin{tabular}{c c c c}
    \toprule
    Processes & Statistical & Theoretical & Experimental \\
    \midrule
    $Z\!\gamma$ production ( $e^+e^- \!\!\to\hsm\ell^+\ell^- \gamma$ ) 
    & 0.52\% & 0.5\% & (+2.96,\,-3.15)\% \\[1mm]  
    \multirow{2}{*}{Fixed background} & \multicolumn{3}{c}{Dominant background: 5\%} \\
                                      & \multicolumn{3}{c}{Other backgrounds: 100\%} \\
    \bottomrule
    \end{tabular}
\caption{\small Overview of systematic uncertainties, estimated for 
$\sqrt{s} \!=\! 240$\,GeV with integrated luminosity of 20\,ab$^{-1}$.\ 
Those in the signal process are separated into statistical, theoretical, 
and experimental categories.\
The signal uncertainty is attributed predominantly to experimental factors, including resolution, identification efficiencies, and detector acceptance, 
collectively termed as ``Experimental".
Background events are floated manually to account for potential uncertainties, 
according to the prescription in~\cite{CEPCZH_BaiYu}.}
\label{tab:syst_table}
\label{tab:6}
\end{table}

%% file: Results.tex
The expected event yields for the SM $Z\!\gamma$ process and backgrounds are 
summarized in Table\,\ref{tab:Eventyields_table}, as obtained after applying all event-topology based selections.\ 
The expected yields of SMEFT samples are propagated to the SMEFT fitting framework 
including all systematic uncertainties, and used to obtain sensitivities for the nTGCs.\

\begin{figure}[!htb]
\centering
\includegraphics[width=.7\columnwidth]{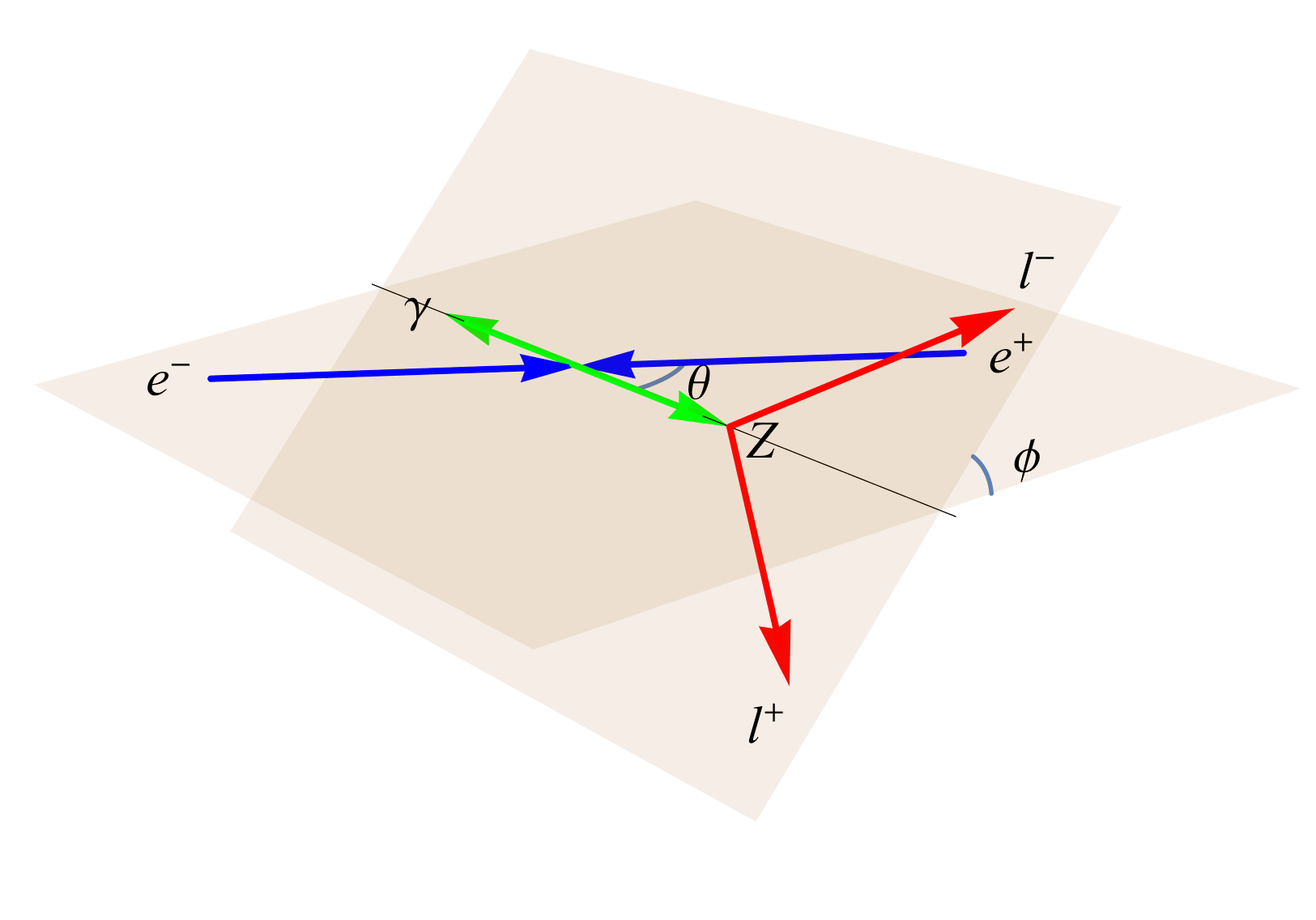}
\vspace*{-4mm}
\caption{\small 
Kinematics in the $e^+e^-$ collision frame of the reaction $e^+e^-\!\!\to\!Z\gamma$
followed by the leptonic decays $Z\!\to\!\ell^+\ell^-$ 
\cite{Ellis_2020}\cite{Ellis_2023_lly}.\
We define $\phi$ as the angle between the scattering plane 
and the decay plane of the $Z$ in the $\ell^+\ell^-$ center-of-mass frame, and
$\theta$ is the polar scattering angle between the directions of the outgoing $Z$ 
and initial state $e^-$.}
\label{fig:phi_angle}
\label{fig:5}
\end{figure}

\begin{table}[!htb]
    \centering
    \begin{tabular}{c c}
    \toprule
    Processes & Event Numbers ($\times\hsm 10^{3}$)  \\
    \midrule
    SM $Z\!\gamma$ production & $3842.7^{+117.1}_{-124.2}$  \\
    2-fermion background      & 5965.1 $\pm$ 298.3    \\
    4-fermion background      & 8.0    $\pm$ 8.0      \\
    Higgs background          & 8.1    $\pm$ 8.1     \\
    \midrule
    Total yield    & $9823.8^{+320.7}_{-323.3}$   \\
    \bottomrule    
    \end{tabular}
    \caption{\small SM event yields and uncertainties ($\times 10^{3}$), extracted at $\sqrt{s} \!=\!240$\,GeV with integrated luminosity of 20\,ab$^{-1}$.\ 
    The expected event yields, incorporating both electron and muon channels, are extracted after applying all event topology-based selections described in the text.\ 
    The estimates include both statistical and systematic uncertainties. }
    \label{tab:Eventyields_table}
\end{table}

A binned profile-likelihood fit is performed to set upper limits on the Wilson coefficients for dimension-8 operators at the 95\% Confidence Level (C.L.). For this purpose we use the EFT fitting framework EFT-fun~\cite{EFTfun} to set 1- and 2-dimensional limits on nTGC parameters, individually and in parameter planes to exhibit their correlations. All the statistical and systematic uncertainties introduced in Section~\ref{sec:systematics} are propagated to the EFT-fun~\cite{EFTfun} framework. 
The kinematic variables $\phi$ and $\theta$ illustrated in Fig.~\ref{fig:phi_angle} are used in this measurement. The interference between SM and pure BSM contributions can be inferred directly from measuring these two variables, which enables better sensitivities for the nTGC coefficients. 

\begin{figure}[!htp]
\centering
\includegraphics[width=0.47\columnwidth]{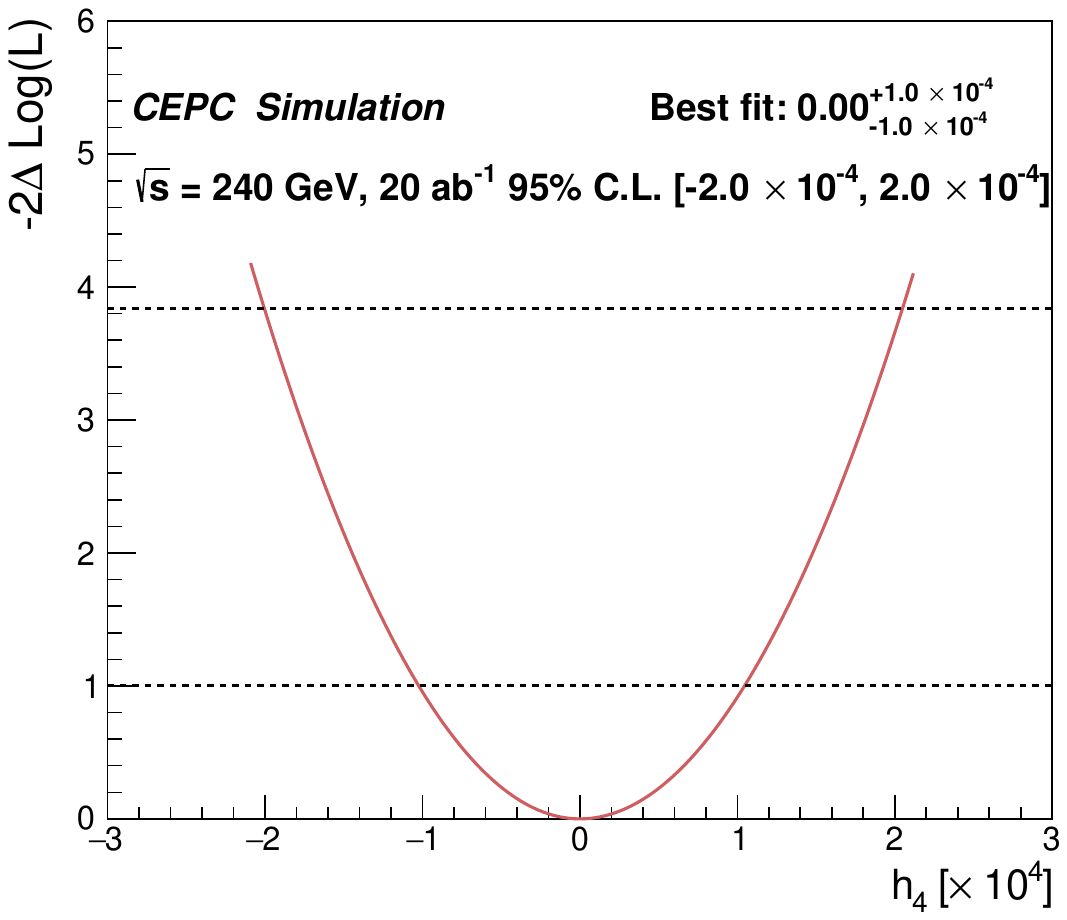}
\includegraphics[width=0.47\columnwidth]{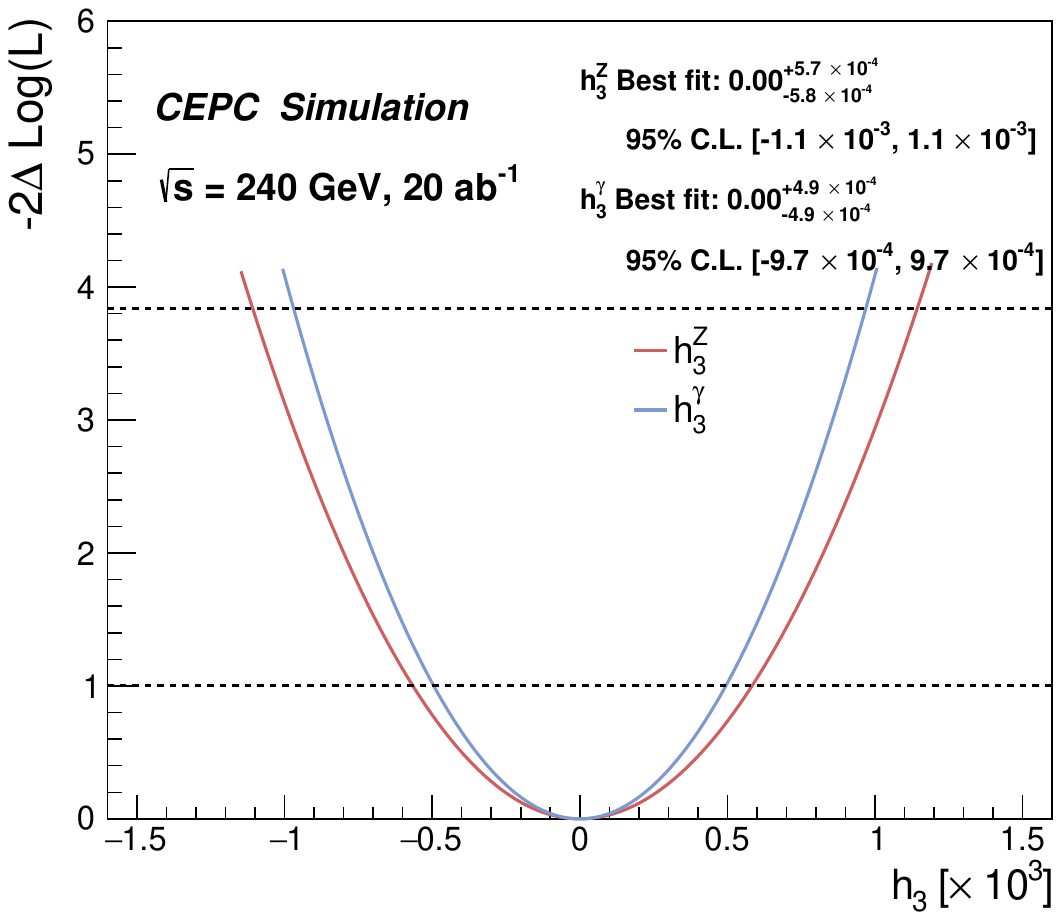}
\vspace*{-3mm}
\caption{\small Expected limits (95\%\,C.L.) on the nTGC form factors 
$(h_{4}^{},\, h_{3}^{Z},\,h_{3}^{\gamma})$ and 1$\sigma$ ranges (dotted lines).\ The best fit values shown in the plots correspond to the best agreements with the SM predictions.}
\label{fig:plot_1D_FF_limits}
\end{figure}

\begin{figure}[!htb]
\centering
\includegraphics[width=0.45\columnwidth]{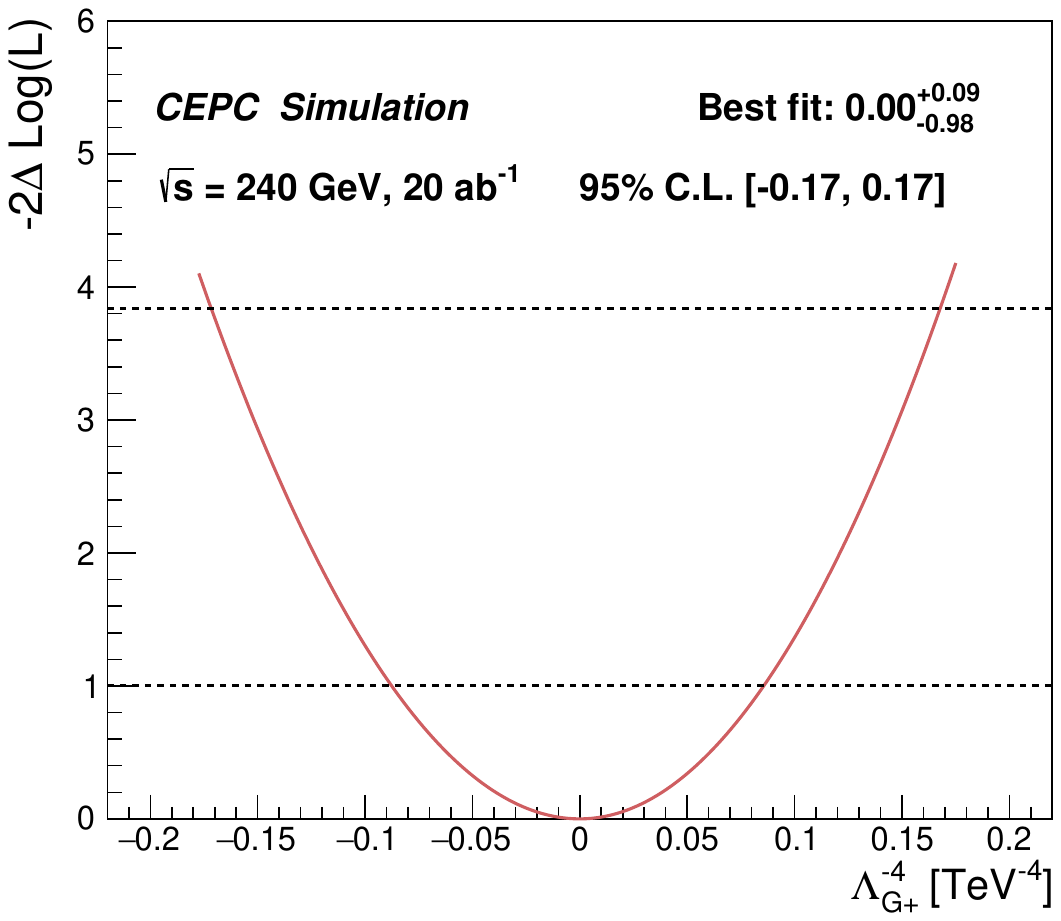}
\includegraphics[width=0.45\columnwidth]{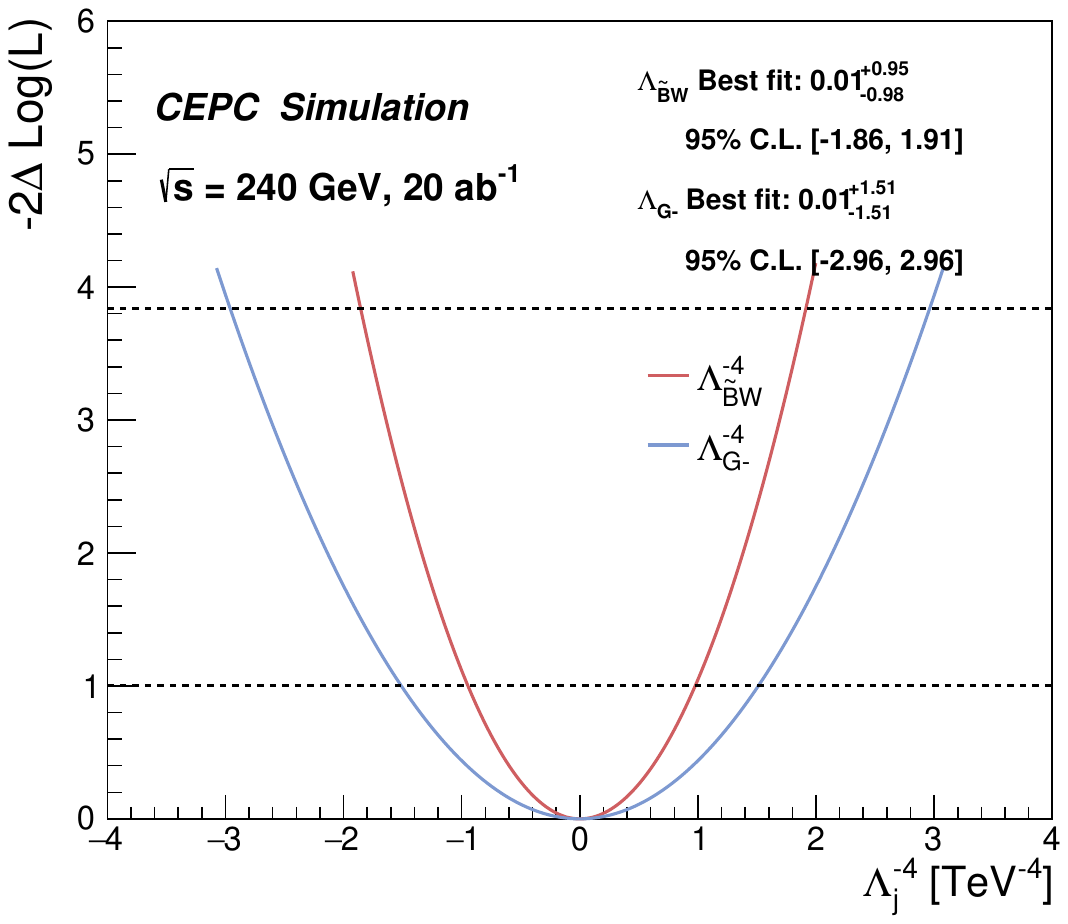}
\vspace*{-3mm}
\caption{\small  Expected limits (95\%\,C.L.) on the coefficients $[\Lambda_j^{-4}]$ (in TeV$^{-4}$) of the dimension-8 nTGC operators 
$(\mathcal O_{G+}^{},\,\mathcal O_{G-}^{},\,\mathcal O_{\tilde{B}W}^{})$ 
and 1$\sigma$ ranges (dotted lines).\
The best fit values shown in the plots correspond to the best agreements with the SM predictions.
} 
\label{fig:plot_1D_limits}
\end{figure}

Table~\ref{tab:Limits_nTGCs} summarises the sensitivity reaches at the 95$\%$ CL for the new physics scales $\Lambda_i$ as obtained from the expected constraints on the associated form factors  derived from the SMEFT dimension-8 coefficients given in the Effective Vertex Approach in Eq.~(\ref{eq:h_conversion}), with all the systematic uncertainties taken into account.
The constraints on the form factors derived in the Effective Vertex Approach are shown in Fig. \ref{fig:plot_1D_FF_limits} and the corresponding constraints on the operator scales within the SMEFT framework are shown in Fig. \ref{fig:plot_1D_limits}. Both figures highlight the central 95 $\%$ C.L range of the integral over the likelihood distribution, while values outside this range are excluded at this level. These depictions of the expected constraints on both the form factors and corresponding dimension-8 operator coefficients within the SMEFT framework offer a comprehensive understanding of the sensitivities to individual higher-dimensional operators.

\begin{table}[!htb]
    \centering
    \begin{tabular}{c c |  c c}
    \toprule
    Form Factors &     Expected limits &  New Physics Scales    
    & Expected limits\,(TeV) \\
    \hline 
 &&& \\[-3.5mm]
    $h_{4}^{}$                             &  [$-2.0,\, 2.0]\!\times\! 10^{-4}$  & $\Lambda_{G+}^{}$\,               &  1.55   \\
    $h_{3}^{\gamma}$                       &  [$-9.7,\, 9.7]\!\times\! 10^{-4}$  & $\Lambda_{G-}^{}$\,               &  0.76   \\
    $ h_{3}^{Z}$                           &  [$-1.1,\, 1.1]\!\times\! 10^{-3}$  & $\Lambda_{\tilde{B}W}^{}$\,       &  0.85   \\
                                           &  & $\Lambda_{\widetilde{BW}}^{}$\,   &  1.05 \\
    \bottomrule
    \end{tabular}
    \caption{\small Sensitivity reaches for the new physics scales
    $\Lambda_i$ and the form factors $(h_4^{},\,h_3^{\gamma},\,h_3^{Z})$ at the 95\%\,C.L.,
    which are obtained by analyzing the $\ell^+\ell^-\gamma$ channels with a benchmark luminosity of 20\,ab$^{-1}$ 
    and collision energy $\sqrt{s\,} \hsm = 240$\,GeV.}
\label{tab:Limits_nTGCs}
\label{tab:8}
\end{table}

\begin{figure}[!ht]
\centering
\includegraphics[width=.42\columnwidth]{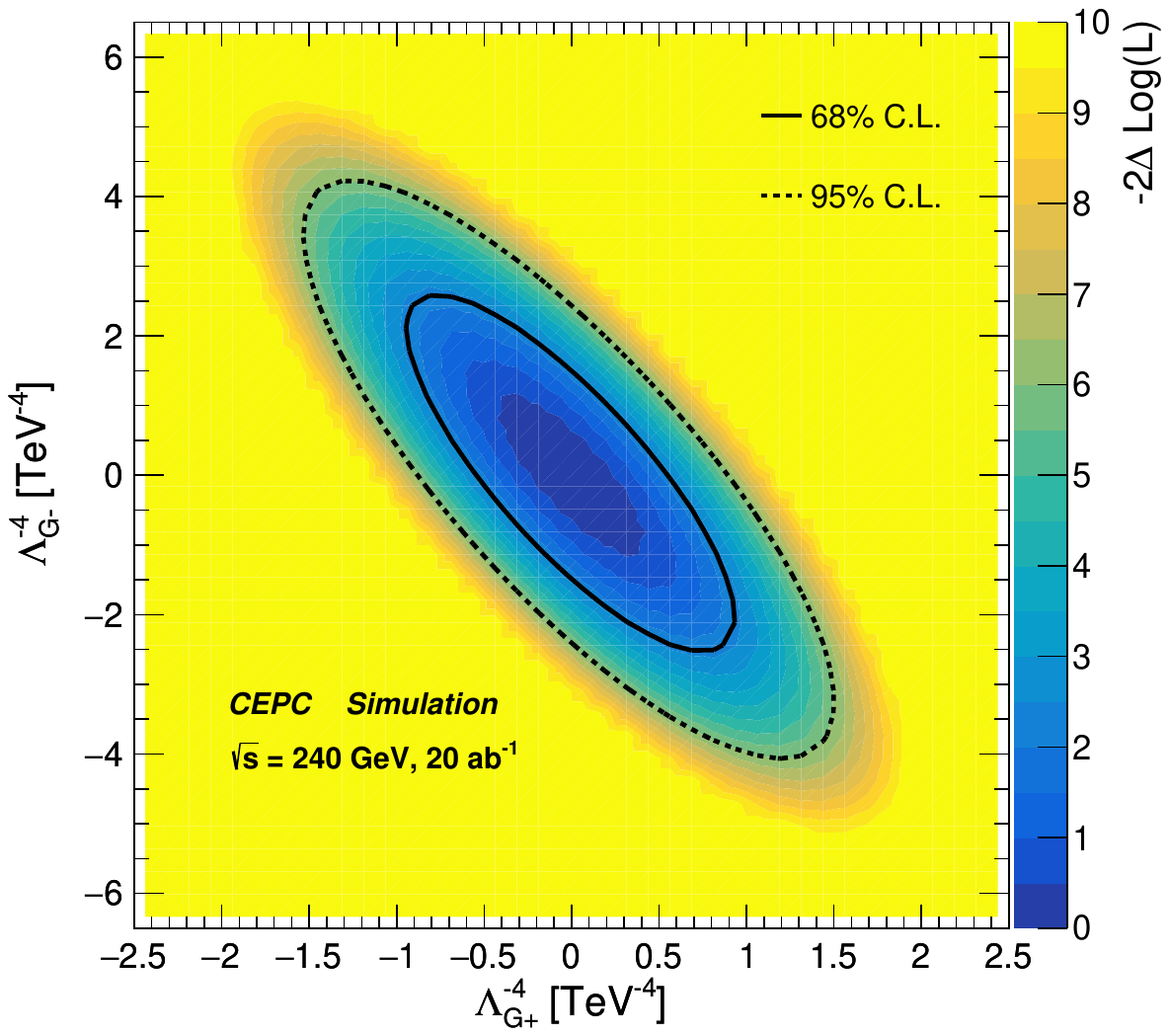}
\includegraphics[width=.42\columnwidth]{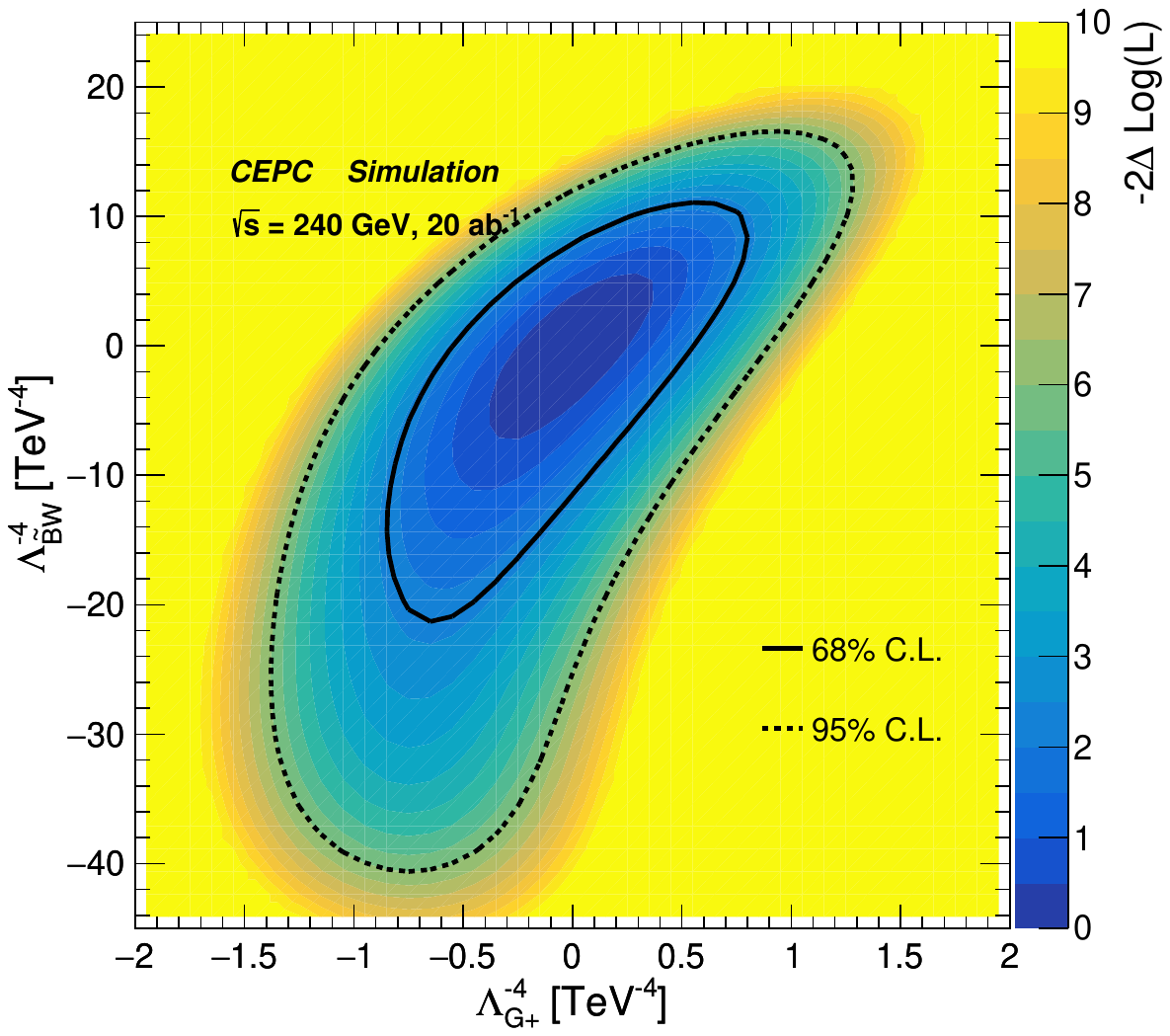}
\caption{\small Correlation contours at the 68\% and 95\%\,C.L. for the cutoff scales 
$\Lambda_j^{}$ of each pair of dimension-8 nTGC operators, where the axis labels are in units of TeV$^{-4}$.} 
\label{fig:ExpectedLimits_Unconverted_2D}
\label{fig:9}
\end{figure}

In adition to these 1-dimensional limits, we have also studied the constraints on different pairs of form factors, so as to understand their allowed correlations. Constraints in 2-dimensional planes are displayed as contour plots in Fig.\,\ref{fig:ExpectedLimits_Unconverted_2D}.\ The solid lines in these plots represent the experimental constraints at 68\%\,C.L., while the dashed lines indicate the 95\%\,C.L. constraints, and areas outside the dashed (approximate) ellipses are excluded 
at the 95\%\,C.L., taking into account all systematic uncertainties.
We observe that the contour plots exhibit significant correlations between pairs of form factors. 

\begin{figure}[!ht]
\centering
\includegraphics[width=.42\columnwidth]{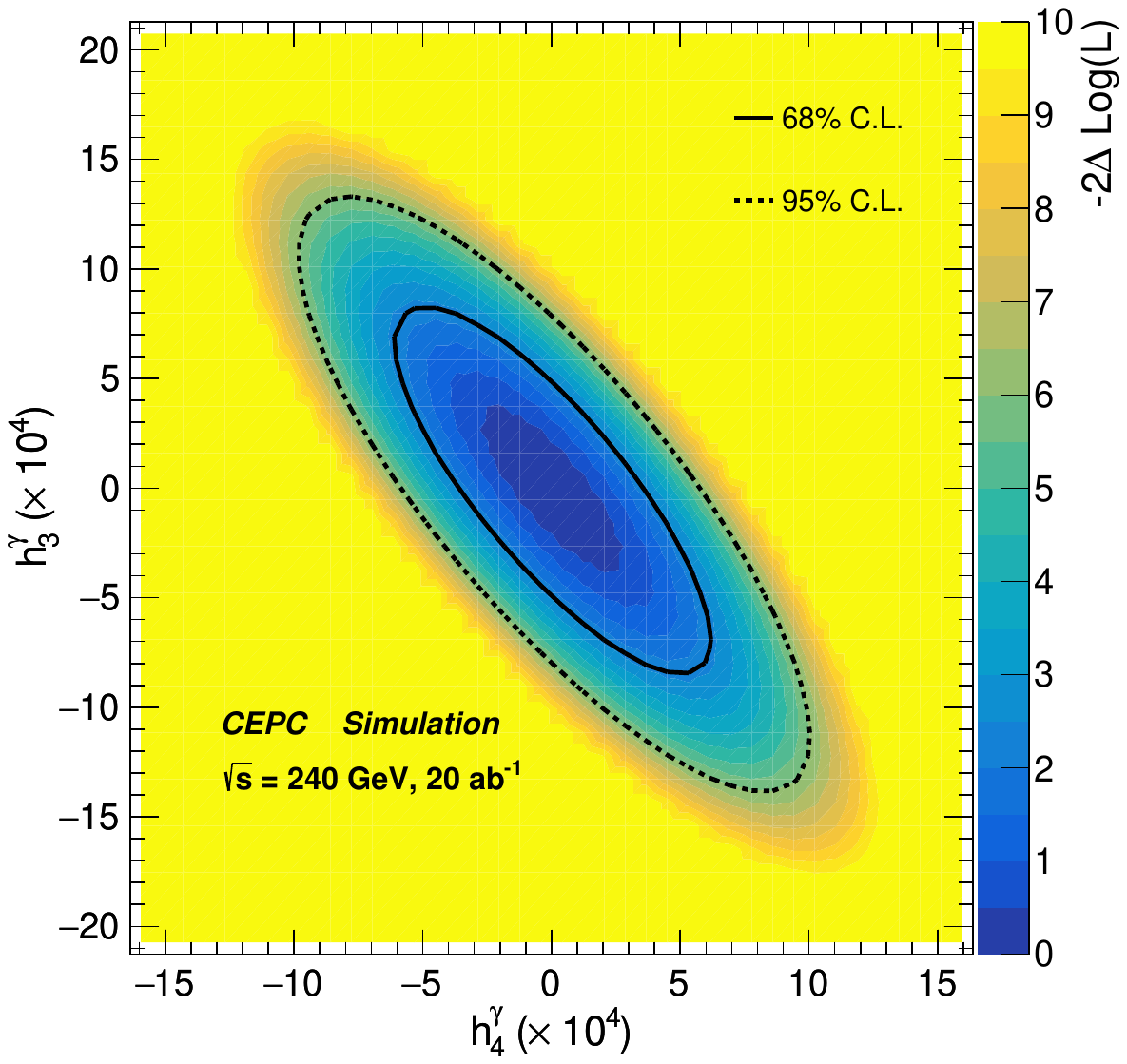}
\includegraphics[width=.42\columnwidth]{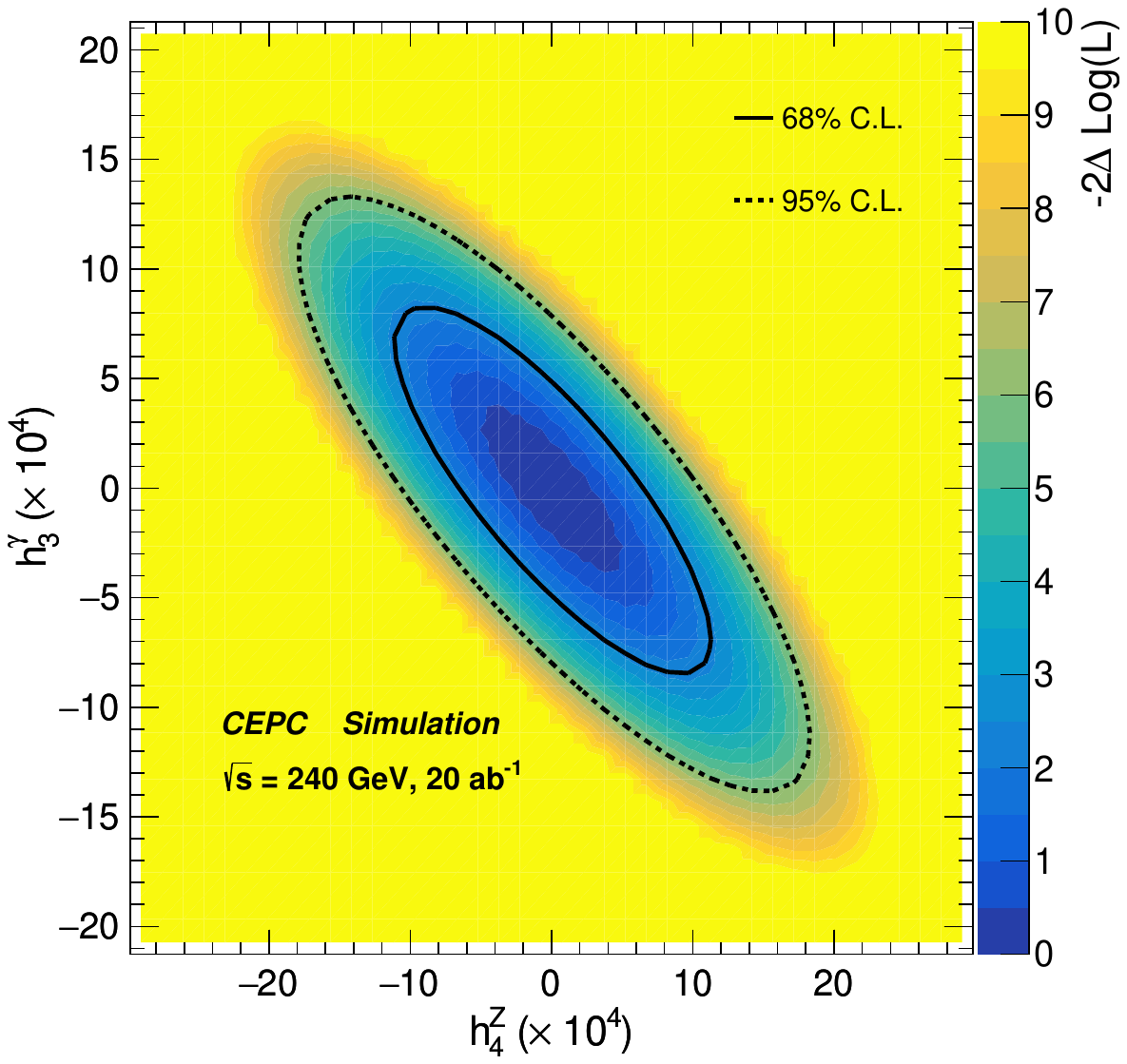} \\
\includegraphics[width=.42\columnwidth]{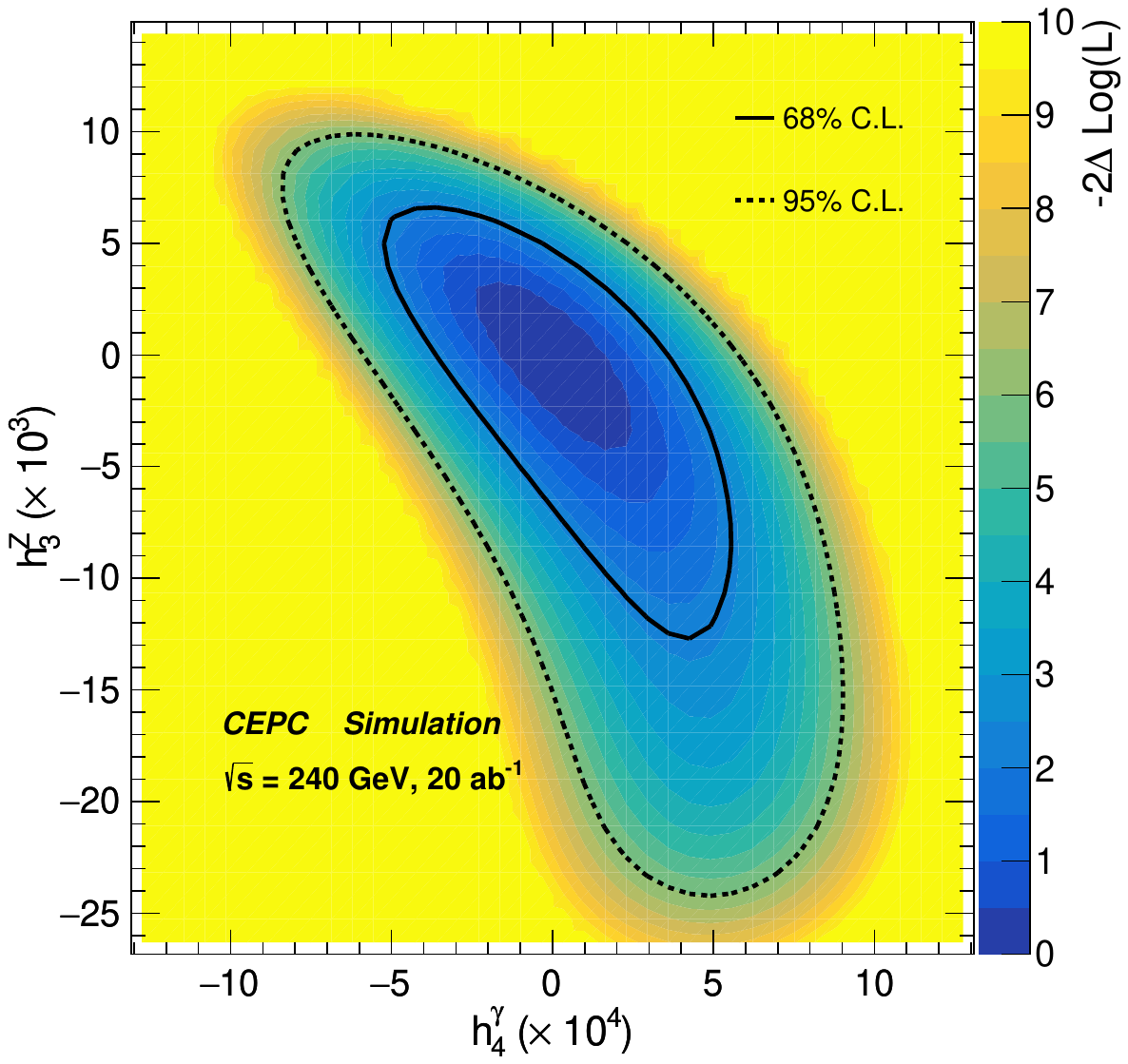}
\includegraphics[width=.42\columnwidth]{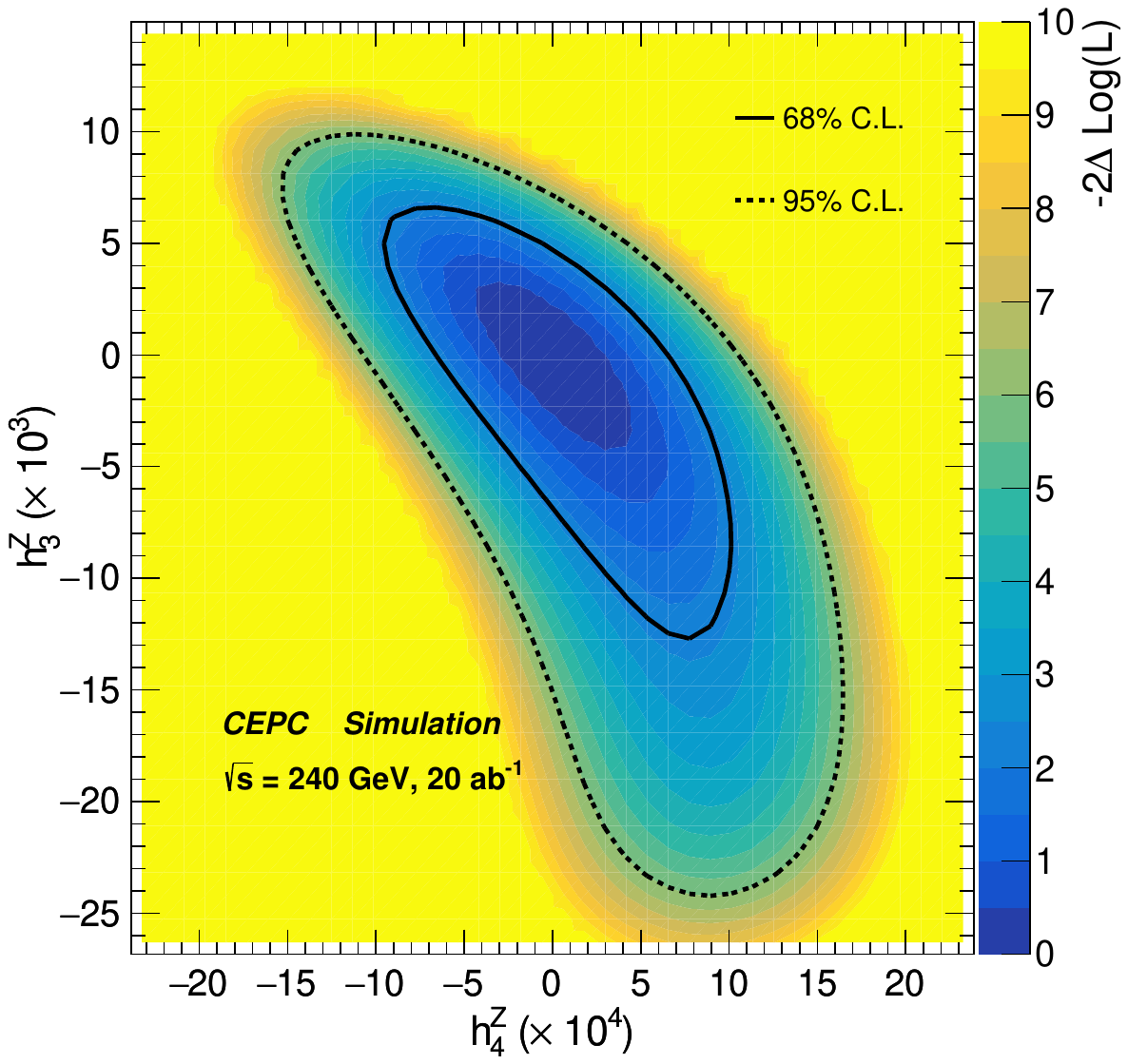} 
\vspace*{-3mm}
\caption{\small Correlation contours at the 68\% and 95\%\,C.L. for each pair of nTGC form factors. }
\label{fig:ExpectedLimits_2D}
\label{fig:8}
\end{figure}

As an alternative visualisation of our results, we have transformed the constraints from this form factor analysis to limits on the scales of the corresponding dimension-8 SMEFT operators in Fig.~\ref{fig:ExpectedLimits_2D}. The aspect ratios and orientations of the (approximately) elliptical contours indicate the degrees of correlation between pairs of operator coefficients. 

The expected limits obtained in this paper are slightly better than the phenomenological results estimated theoretically in~\cite{NewEFTnTGC}, despite the inclusion of all sources of systematic uncertainties, such as detector acceptance, object reconstruction, identification efficiencies and resolution. 
In this measurement, we defined a new variable $u = \cos\theta \times \cos\theta^*$ (where $\theta^*$ is the decay angle as measured in the Z boson's rest frame).
By employing the BDTG method, we constructed a decision boundary in the $\phi - u$ plane to distinguish efficiently between events with positive and negative cross sections. 
This approach represents an improvement over the theoretical method that used only a single variable ($\phi$), also achieves better significance and accuracy.
This method is documented in Appendix \ref{sec:appendix_BDTG}.

%% file: Conclusions.tex

Since nTGC vertices $Z\gamma V^*$ do not arise in the dimension-4 SM Lagrangian or in the SMEFT at the dimension-6 level, probing them from the contributions of dimension-8 
operators provides a unique opportunity to explore new physics 
beyond the Standard Model (SM).\ 
We have investigated in this work the sensitivities to nTGCs 
through the reaction $e^+ e^- \!\!\to\! \ell^+ \ell^- \gamma$ 
(with $\ell \!=\! e , \mu$), performing a detector-level analysis and simulation for an experiment at the CEPC. Experiments at other $e^+e^-$ colliders with similar integrated luminosities and collision energies are expected to have similar sensitivities
for probing the nTGCs.

Previous studies of the nTGC vertex $Z\gamma V^*$ via form factors are not consistent 
with spontaneously-broken electroweak gauge symmetry of the SM.\  
Recently a new formulation of the nTGC form factor framework 
has been proposed\,\cite{Ellis_2023_lly,Ellis_2023_nny}, 
which is consistently determined by mapping to the complete set of  
dimension-8 nTGC operators of the SMEFT and hence is compatible with  
the full electroweak gauge symmetry of the SM.\  
It was found that extra dimension-8 nTGC operators are needed 
to establish the consistent mapping from 
the dimension-8 nTGC operators to the correct nTGC form factors\,\cite{Ellis_2023_lly,Ellis_2023_nny}.
The consistent form factor expression 
for the CP-conserving nTGC vertex $Z\gamma V^*$ is shown in
Eq.(\ref{eq:ZyVertex_Expression_02}).\ 

We have adopted the new nTGC form factor formula (\ref{eq:ZyVertex_Expression_02})
to analyze the sensitivities to nTGCs in the $Z\gamma$ channel 
with $Z$ leptonic decays based on the benchmark luminosity 20\,ab$^{-1}$  
and $e^+e^-$ collision energy 
$\sqrt{s}\hsm =\hsm 240$\,GeV at the CEPC.\  
With these, we have obtained the nTGC sensitivity limits (95\%\,C.L.) 
that take into account a single nonzero nTGC parameter  
at a time (as shown in Table\,\ref{tab:8}), 
as well as the sensitivity contours (95\% C.L.)  
for each pair of nTGC form factors or for each pair 
of cutoff scales of dimension-8 nTGC operators
(as shown in Figs.\,\ref{fig:8} and \ref{fig:9}).

Our results were obtained by a dedicated simulation with a realistic detector configuration and a full treatment of the systematic experimental uncertainties as well as statistical uncertainties.\ 
A cut-based method is employed for the entire analysis, providing significantly stronger sensitivities compared to previous theoretical analyses~\cite{NewEFTnTGC}.
Additionally, we optimized the extraction of the interference term using the BDTG method, which effectively separates positive and negative events, thereby enhancing the overall sensitivity.

Table\,\ref{tab:Limits_nTGCs} shows that measurements of nTGCs at CEPC and other $e^+e^-$ Higgs factories have the potential to probe energy scales well beyond their center-of-mass energies, even exceeding a TeV in the most sensitive case of the 
nTGC operator ${\cal O}_{G+}^{}$.\ These results are encouraging and confirm 
that nTGC measurements provide an interesting window to the dimension-8 new physics, 
extending the utility of the SMEFT beyond the dimension-6 level.

%% file: Appendix.tex
\renewcommand{\thesubsection}{\thesection.\arabic{subsection}}

\section{Background Samples}
\label{sec:appendix_backgrounds}

We summarize in Table\,\ref{tab:mcsample_backgrounds}
the cross sections of the background samples used in this analysis.\ 
We classify the background samples into 3 categories: 
2 fermions, 4 fermions, and Higgs backgrounds.\ 
Each category contains multiple final states 
and the corresponding cross sections for the different channels 
as presented in this table. 

\begin{table}[h]
    \centering
    \begin{tabular}{c c c c}
        \toprule
        \multicolumn{2}{l}{\hspace*{16mm}Processes} &  Final States & $\sigma$ (fb) \\
        \midrule
        \multirow{3}{*}{2 fermions} & $\ell\ell$        & $e^+e^-/\mu^+\mu^-/\tau^+\tau^-$ & 34856.50 \\
                                    & $\nu\nu$  & $\nu_{e}\bar{\nu}_{e}/\nu_{\mu}\bar{\nu}_{\mu}/\nu_{\tau}\bar{\nu}_{\tau}$    &   50499.51    \\
                                    & $qq$        & $u\bar{u}/d\bar{d}/c\bar{c}/s\bar{s}/b\bar{b}$    &   54106.86    \\
        \midrule
        \multirow{6}{*}{4 fermions} & $WW$\,(hadronic decay) &  & 3825.46   \\
                                    & $WW$\,(leptonic decay) &  & 403.66    \\
                                    & $WW$\,(semi-leptonic decay) &  & 4846.99  \\
                                    & $ZZ$\,(hadronic decay) &  & 516.67    \\
                                    & $ZZ$\,(leptonic decay) &  & 67.81     \\
                                    & $ZZ$\,(semi-leptonic decay) &  & 556.59   \\
        \midrule
        \multirow{5}{*}{Higgs}      & $e^+e^-H$              & $e^+e^- \!+\! H$     &   7.04    \\
                                    & $\mu^+\mu^-H$          & $\mu^+\mu^-\!+\! H$ &    6.77   \\
                                    & $\tau^+\tau^-H$        & $\tau^+\tau^-\!+\! H$ &  6.75   \\
                                    & $\nu\nu H$             & $\nu_{e}\bar{\nu}_{e}/\nu_{\mu}\bar{\nu}_{\mu}/\nu_{\tau}\bar{\nu}_{\tau}\!+\!H$   & 46.29 \\
                                    & $qqH$                  & $u\bar{u}/d\bar{d}/c\bar{c}/s\bar{s}/b\bar{b}\!+\!H$   &   136.81  \\
        \bottomrule
\end{tabular}
\caption{\small Background samples used in the analysis of $e^+ e^- \!\!\to\! Z\gamma$
with the collision energy $\sqrt{s\,}\!=\!240$\,GeV.\  
The background samples are categorised into 3 groups: 
2 fermions, 4 fermions, and Higgs backgrounds.\ Each group including multiple final states 
and the corresponding cross sections for the different channels are summarized in this table.}
\label{tab:mcsample_backgrounds}
\label{tab:9}
\end{table}


\section{Optimisation with kinematic distributions}
\label{sec:appendix_kinematics}
\label{sec:appendix_BDTG}

\begin{figure}[!ht]
    \centering
    \includegraphics[width=.32\columnwidth]{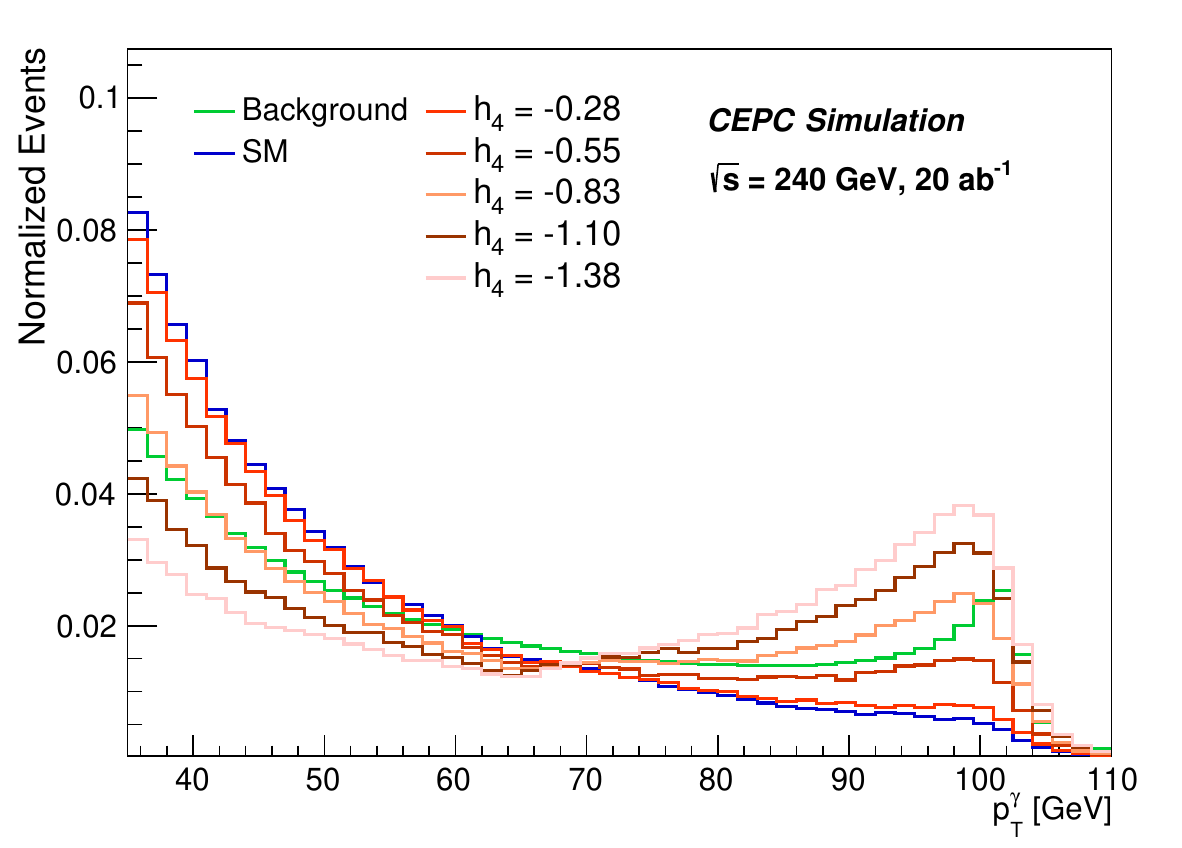}
    \includegraphics[width=.32\columnwidth]{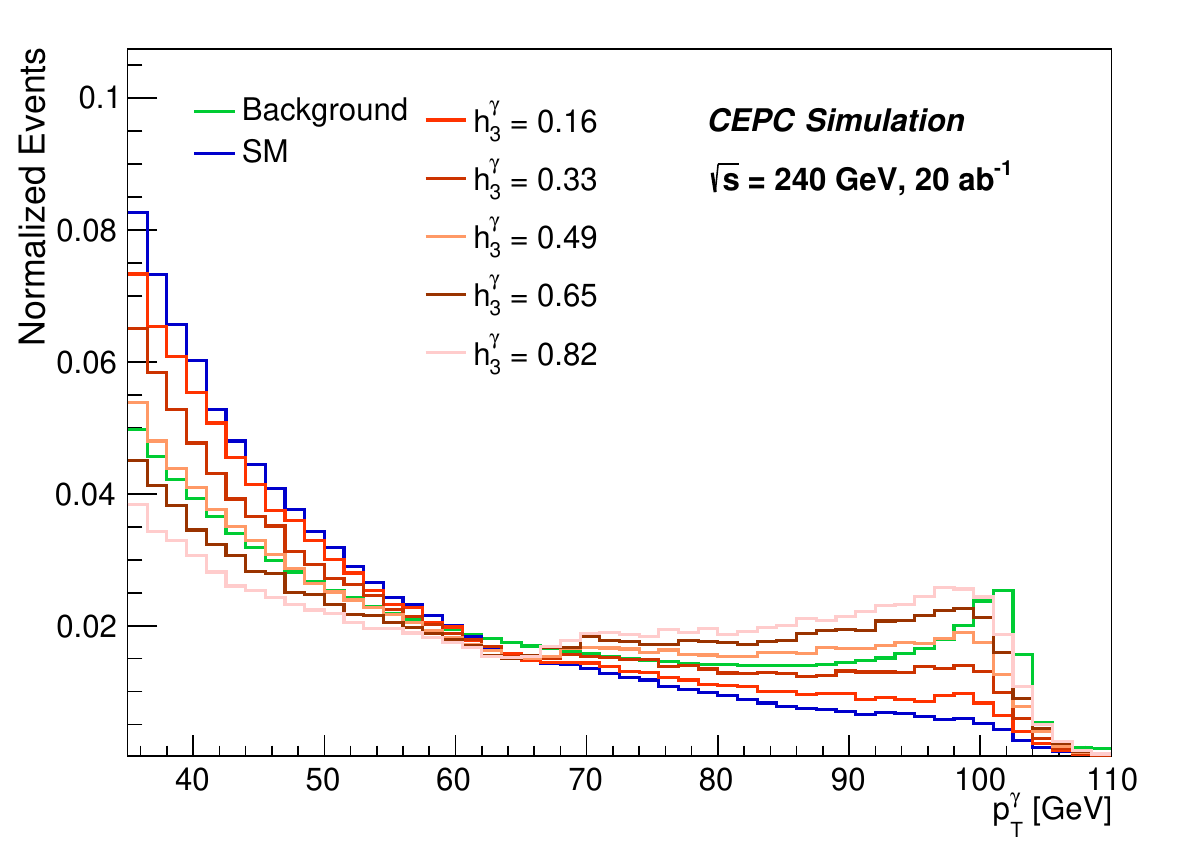}
    \includegraphics[width=.32\columnwidth]{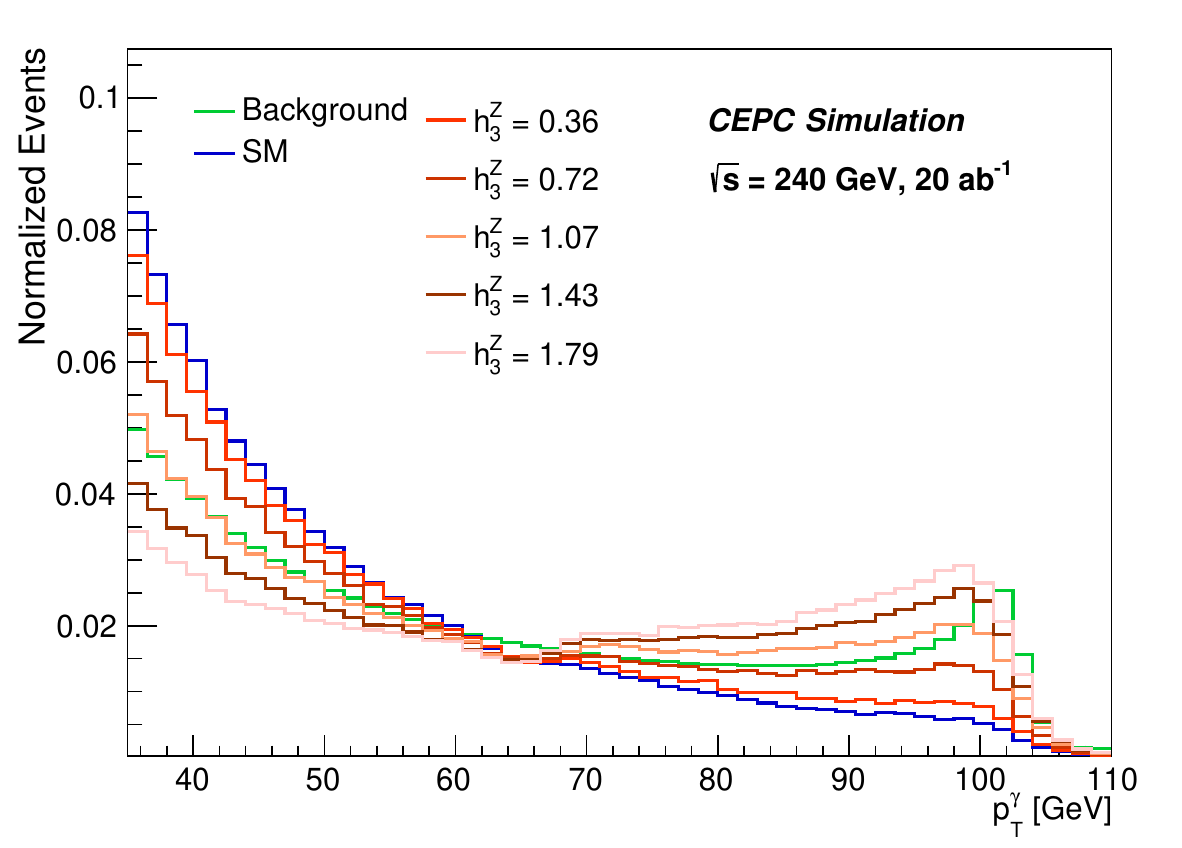}   \\

    \includegraphics[width=.32\columnwidth]{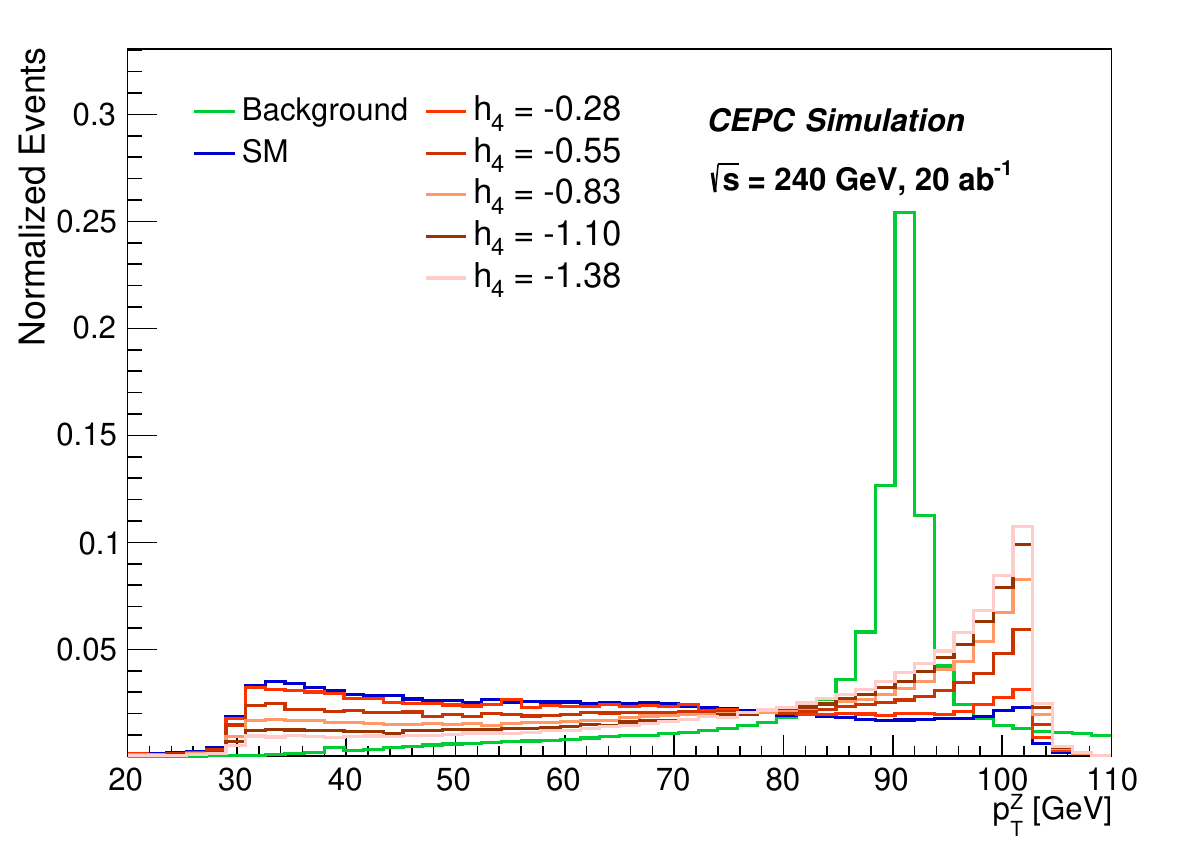}
    \includegraphics[width=.32\columnwidth]{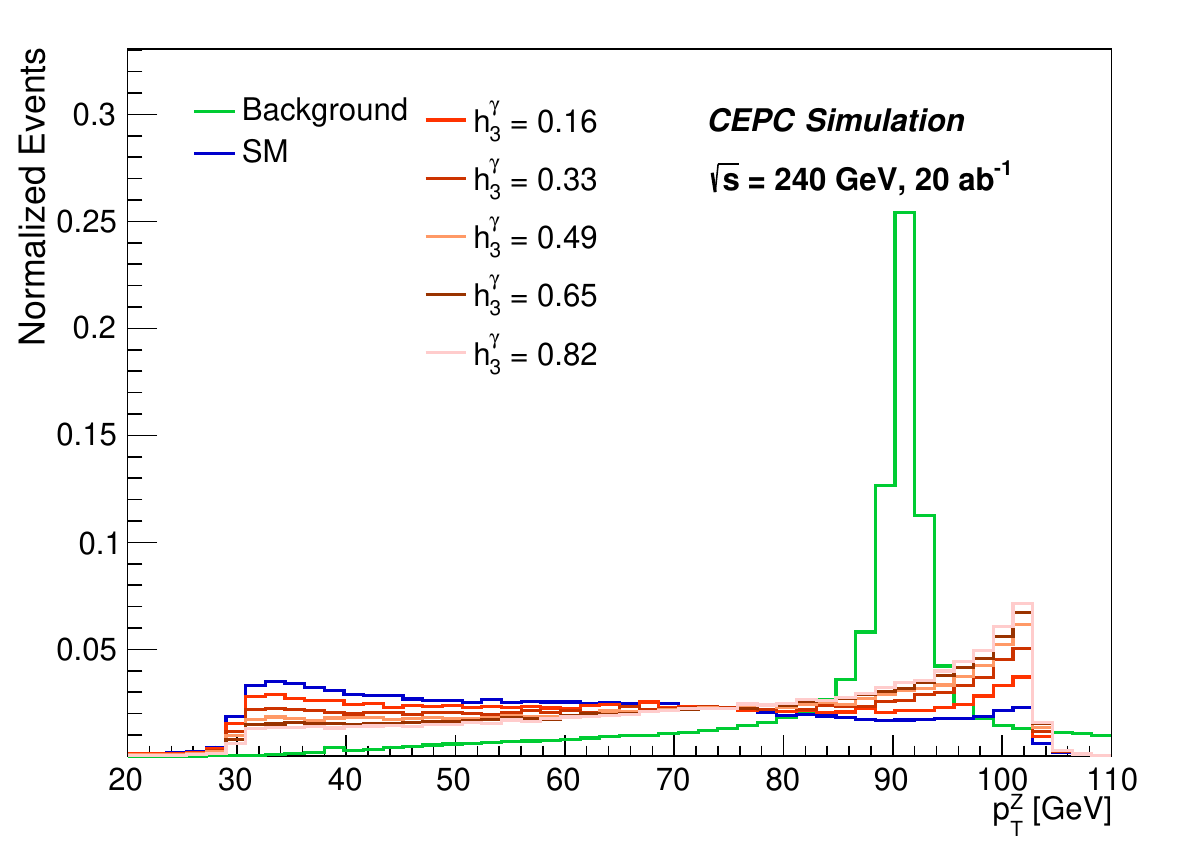}
    \includegraphics[width=.32\columnwidth]{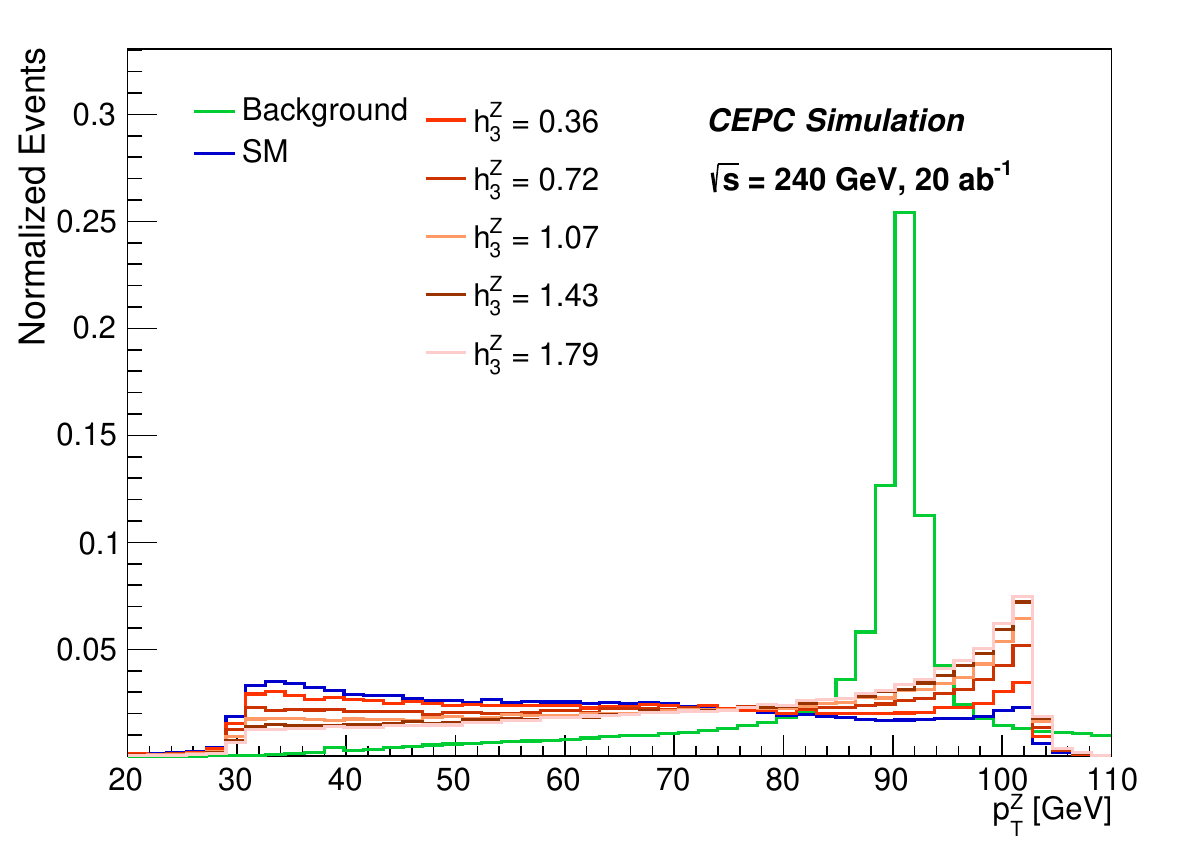}   \\

    \includegraphics[width=.32\columnwidth]{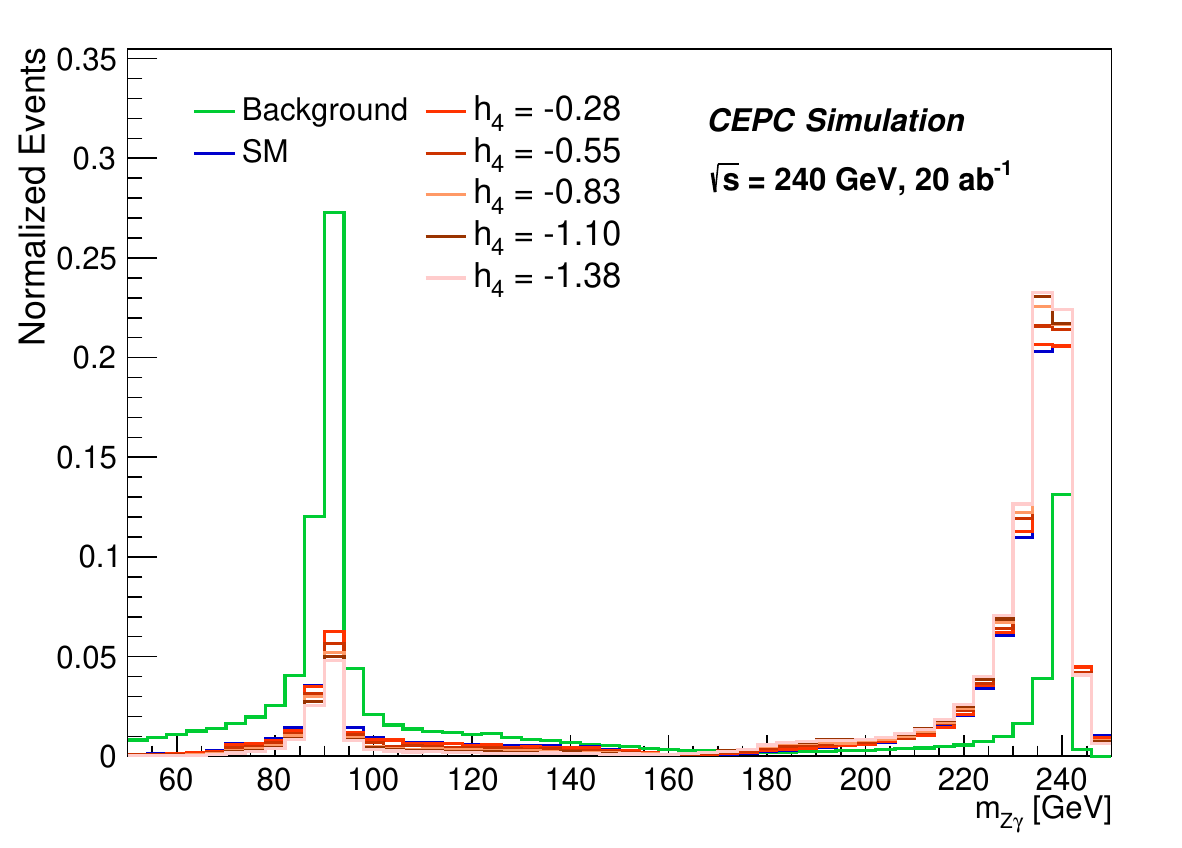}
    \includegraphics[width=.32\columnwidth]{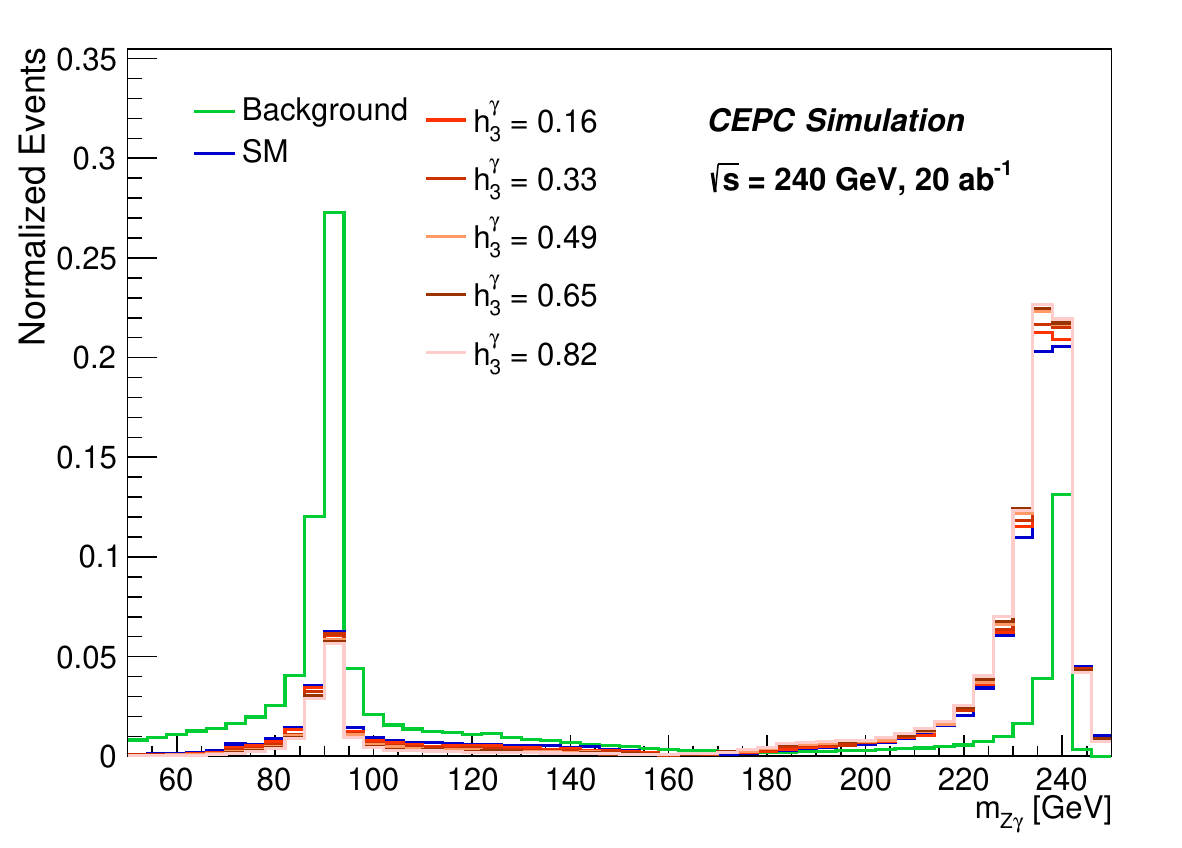}
    \includegraphics[width=.32\columnwidth]{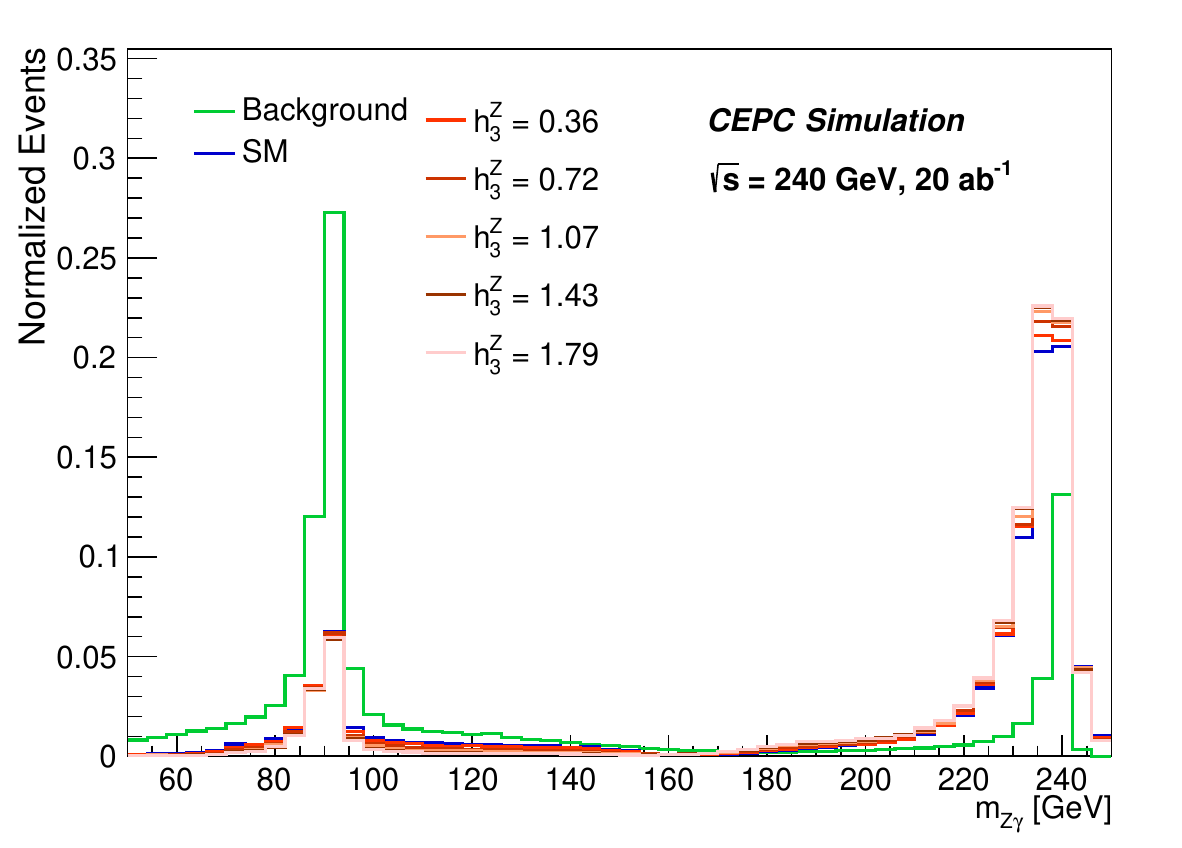}   \\

    \includegraphics[width=.32\columnwidth]{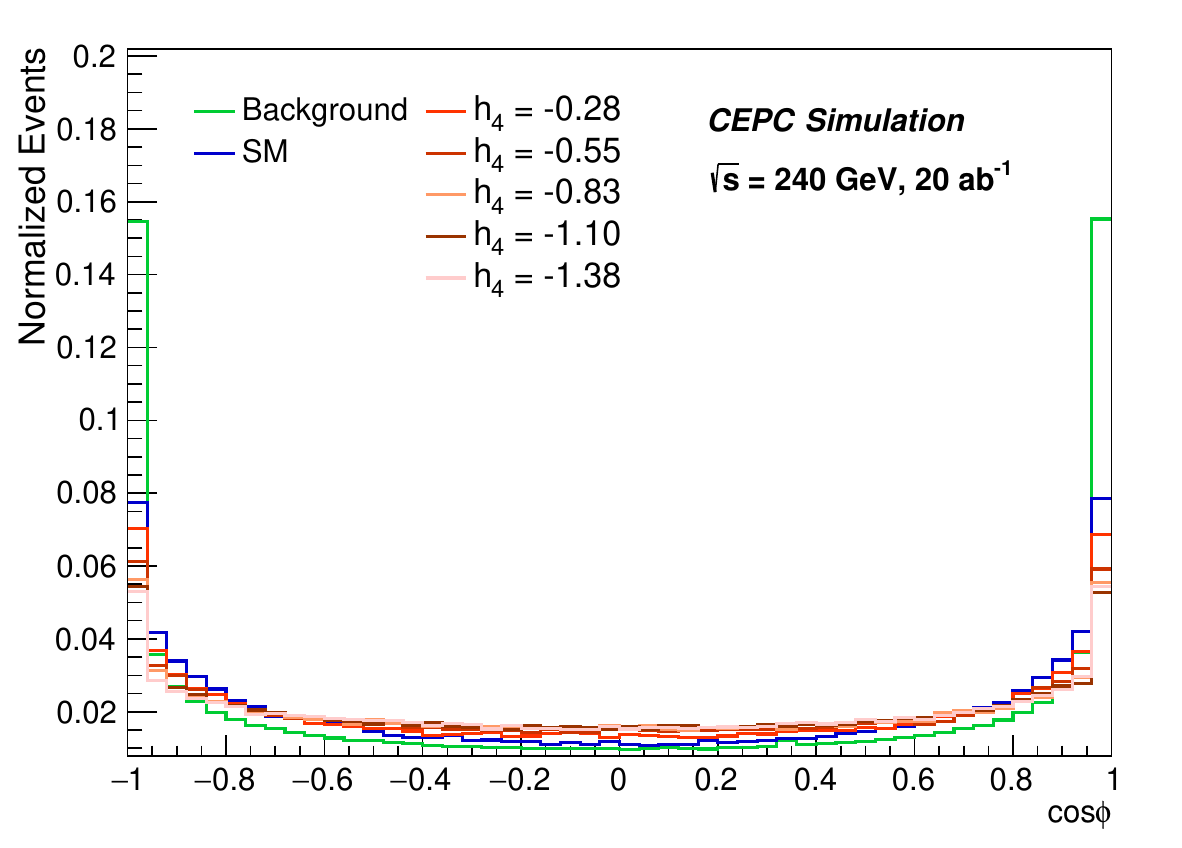}
    \includegraphics[width=.32\columnwidth]{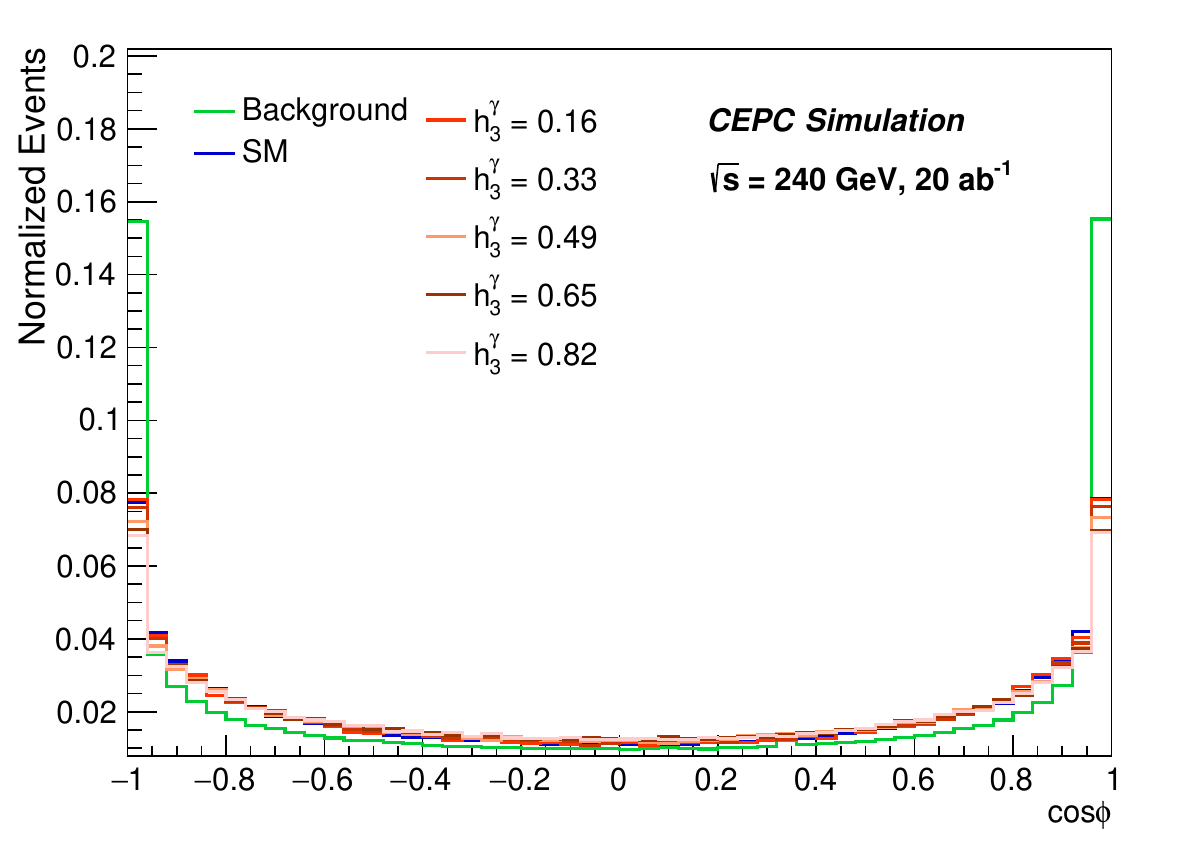}
    \includegraphics[width=.32\columnwidth]{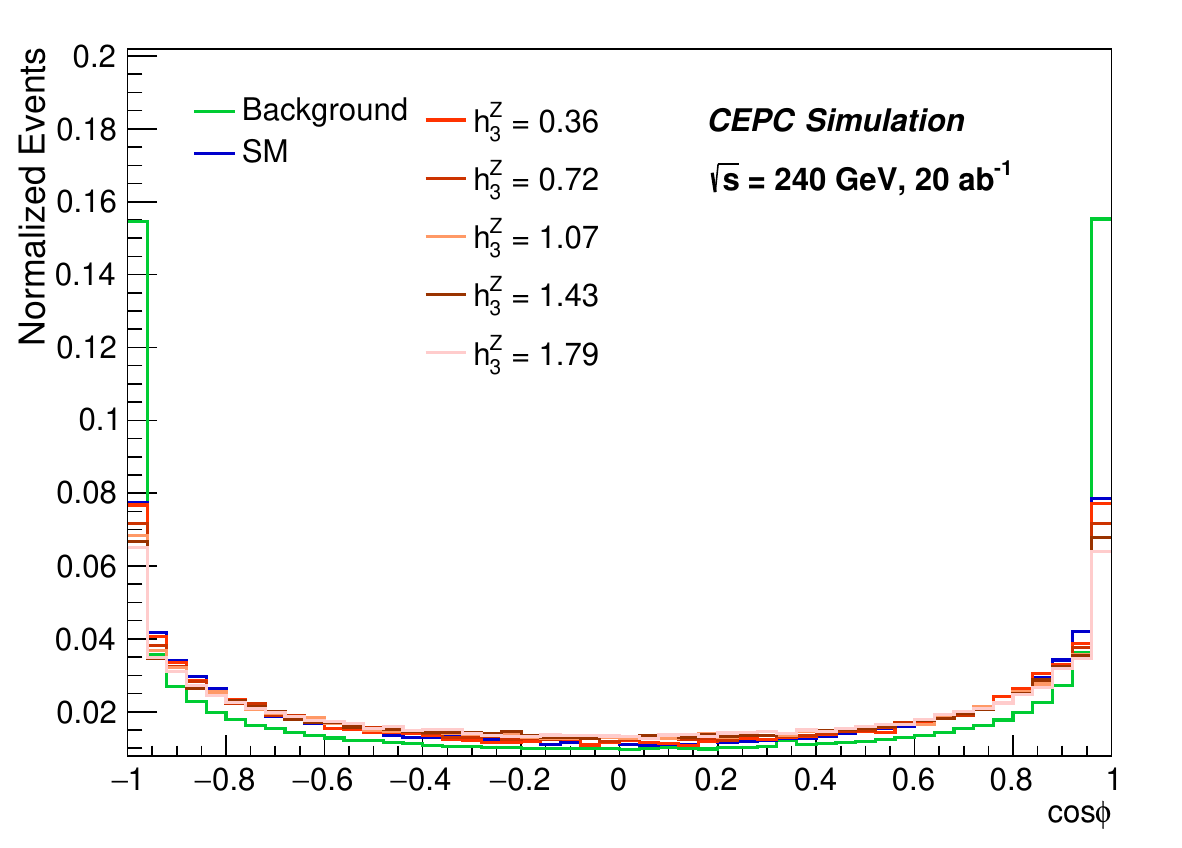}    \\

    \includegraphics[width=.32\columnwidth]{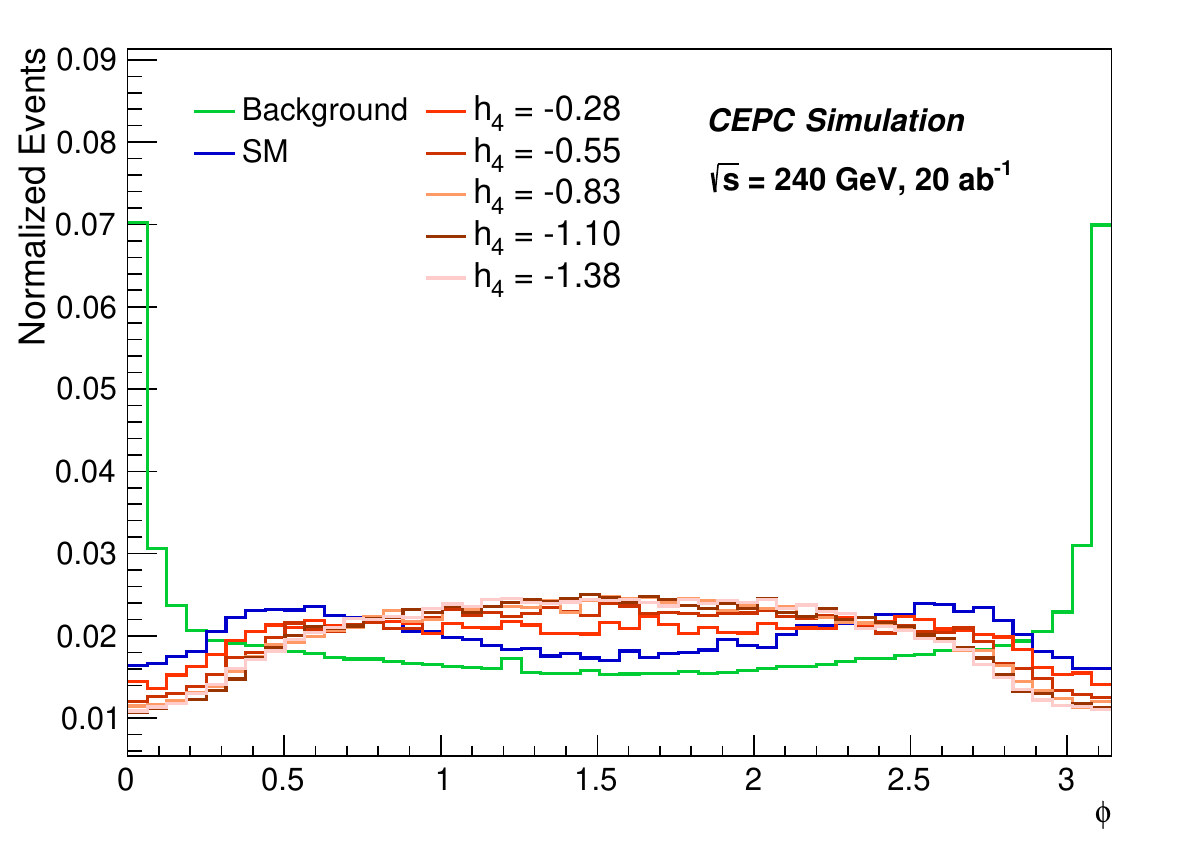}
    \includegraphics[width=.32\columnwidth]{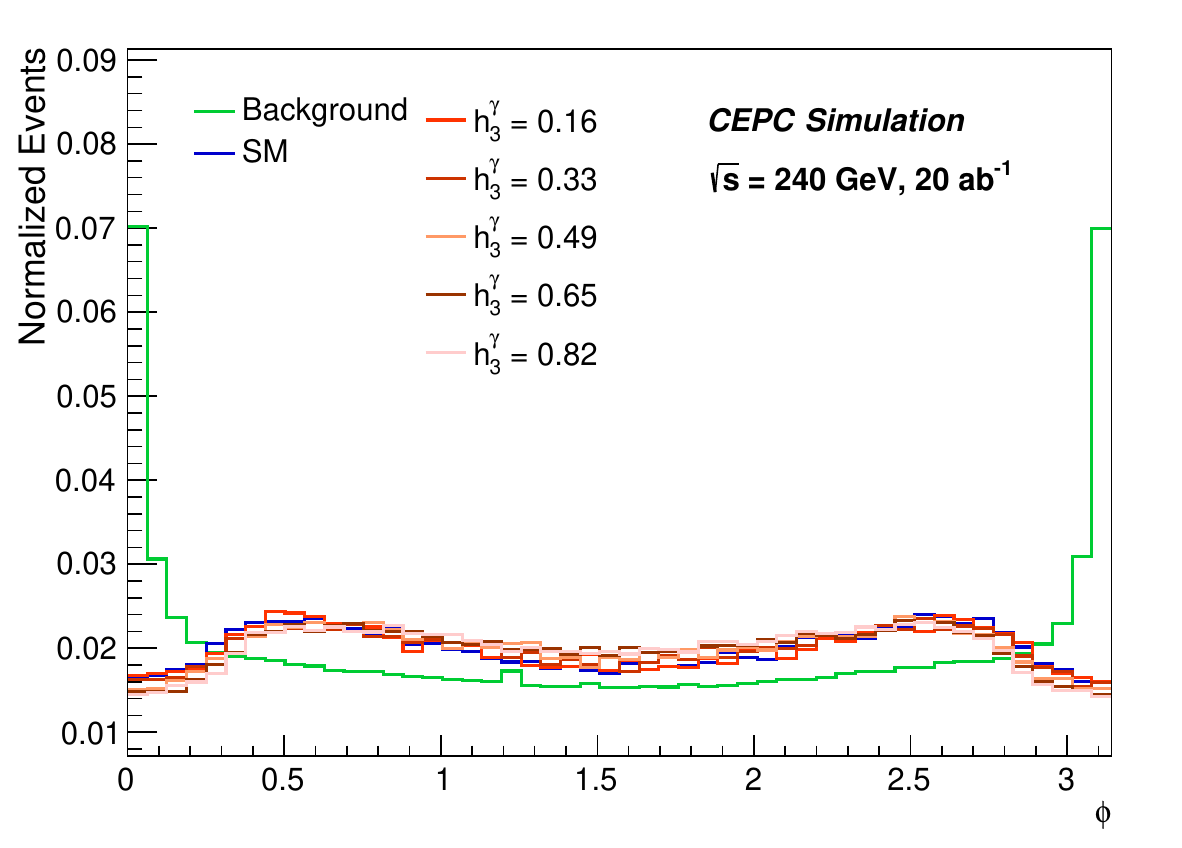}
    \includegraphics[width=.32\columnwidth]{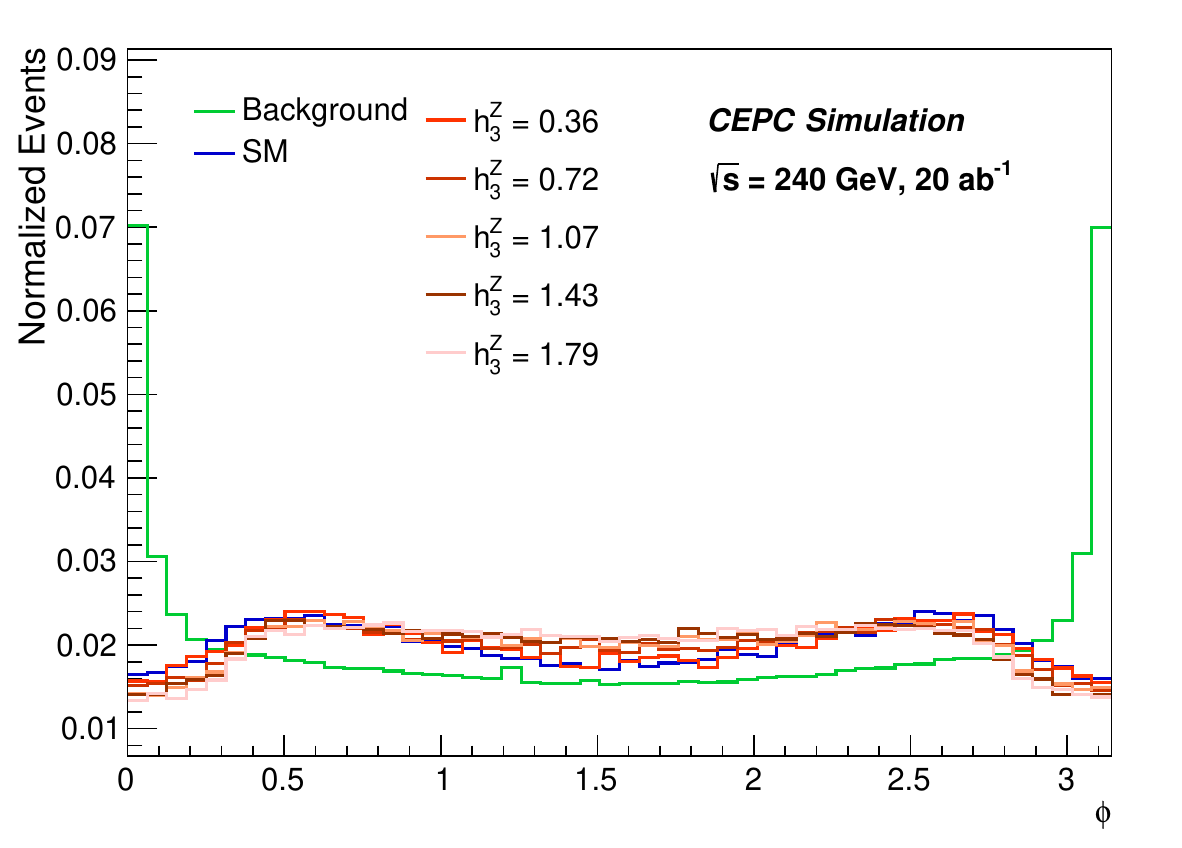}    
    \\[-2mm]
\caption{\small Distributions for kinematic variables and comparisons with the SM backgrounds for different signal processes.\ 
    Samples including 2- and 4-fermion backgrounds and Higgs processes are compared with signal samples generated with varying 
    values of the nTGC form factors.\ Clear differences between the SM $Z\gamma$, SM backgrounds and nTGC $Z\gamma$ processes are visible. }
    \label{fig:kinematic_distribution_multivar}
\end{figure}

We compare the distributions of multiple kinematic variables from different processes 
and display them in the plots of Fig.\,\ref{fig:kinematic_distribution_multivar}.\ 
Differences between the SM $Z\gamma$ process, SM backgrounds and nTGC $Z\gamma$ 
processes (with various form factors) are shown clearly.

\begin{figure}[!ht]
    \centering
    \subfloat[$\bar {\sigma}_1 (h_{4}) >$ 0]{\includegraphics[width=.48\columnwidth]{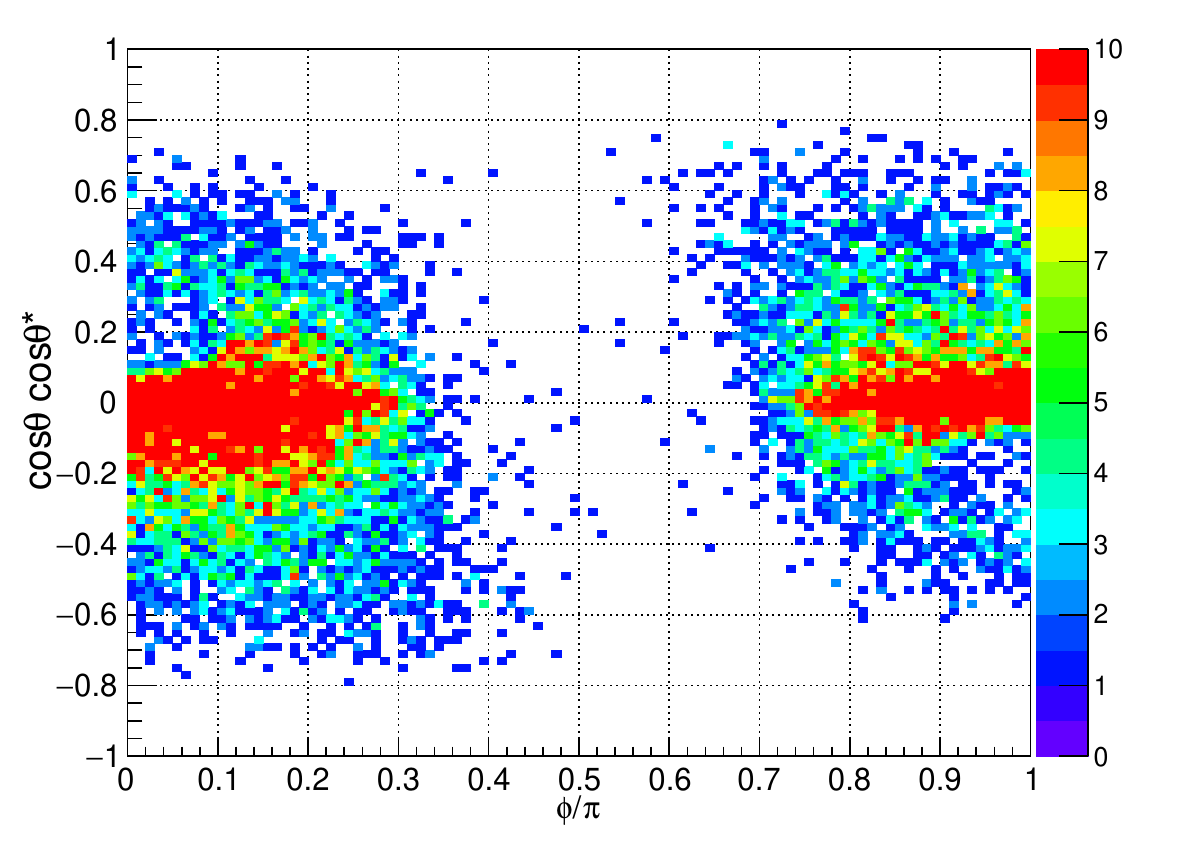}}   
    \subfloat[$\bar {\sigma}_1 (h_{4}) <$ 0]{\includegraphics[width=.48\columnwidth]{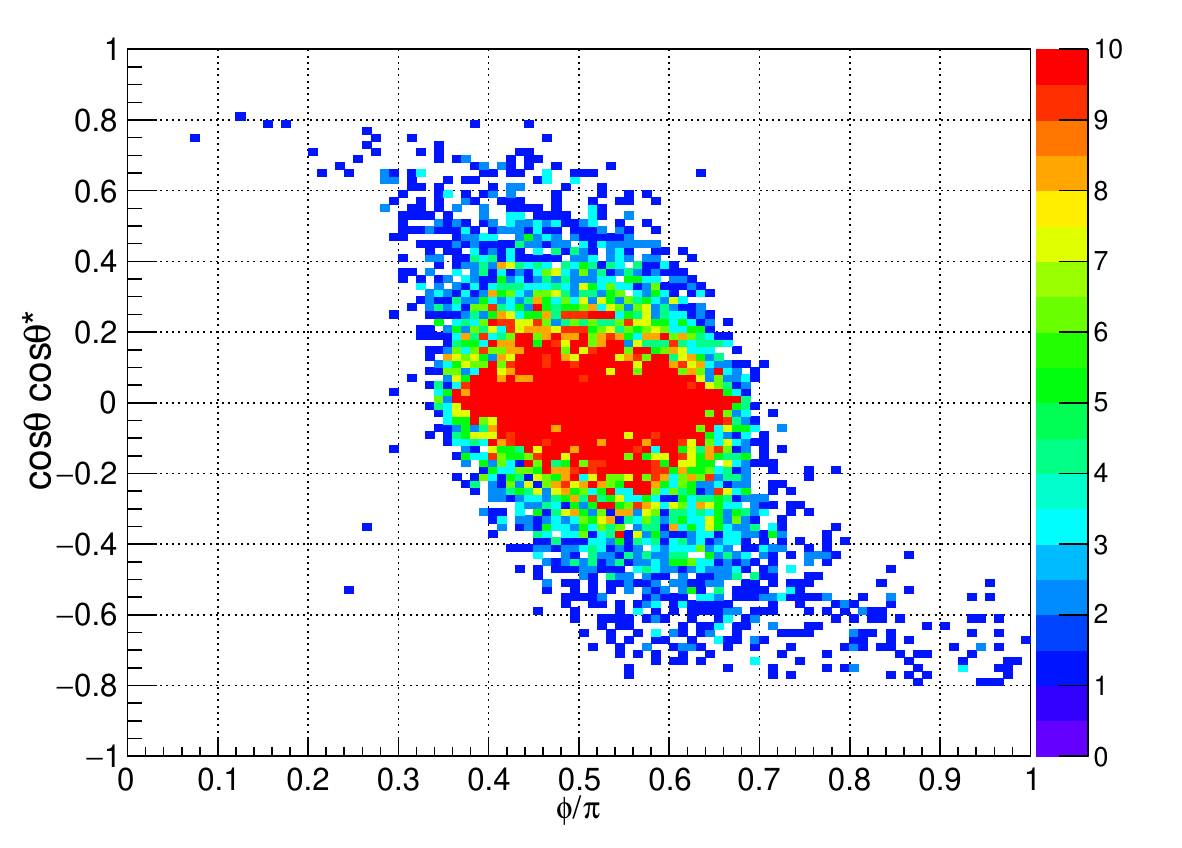}}     \\
    \subfloat[$\bar {\sigma}_1 (h_{3}^{\gamma}) >$ 0]{\includegraphics[width=.48\columnwidth]{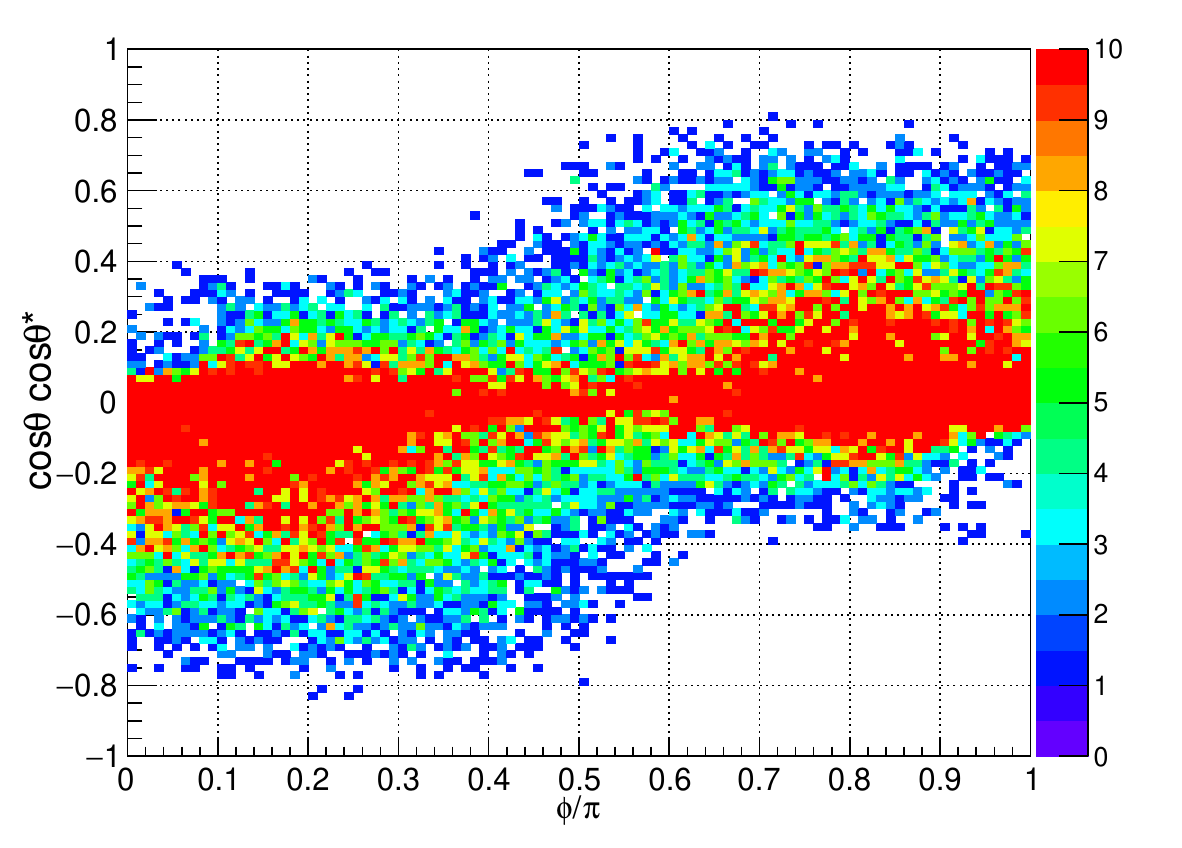}}
    \subfloat[$\bar {\sigma}_1 (h_{3}^{\gamma}) <$ 0]{\includegraphics[width=.48\columnwidth]{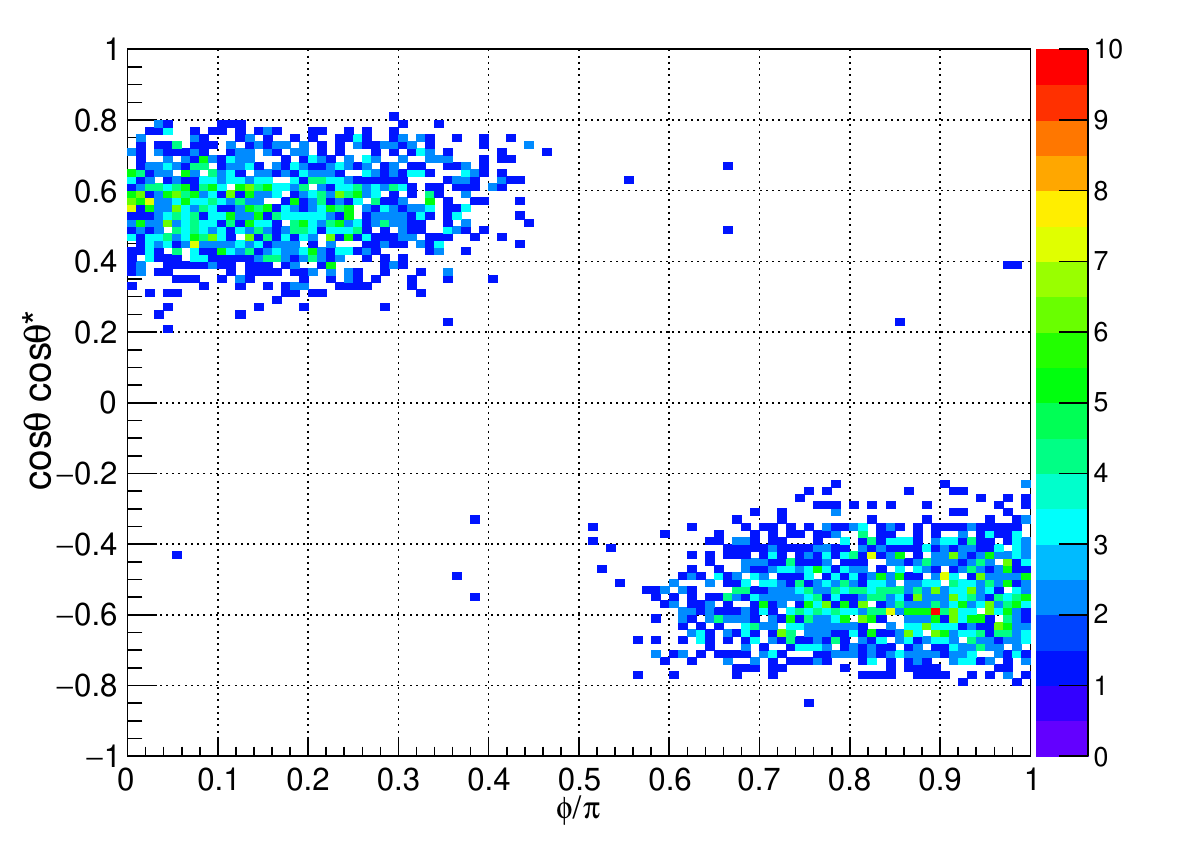}}  \\
    \subfloat[$\bar {\sigma}_1 (h_{3}^{Z}) >$ 0]{\includegraphics[width=.48\columnwidth]{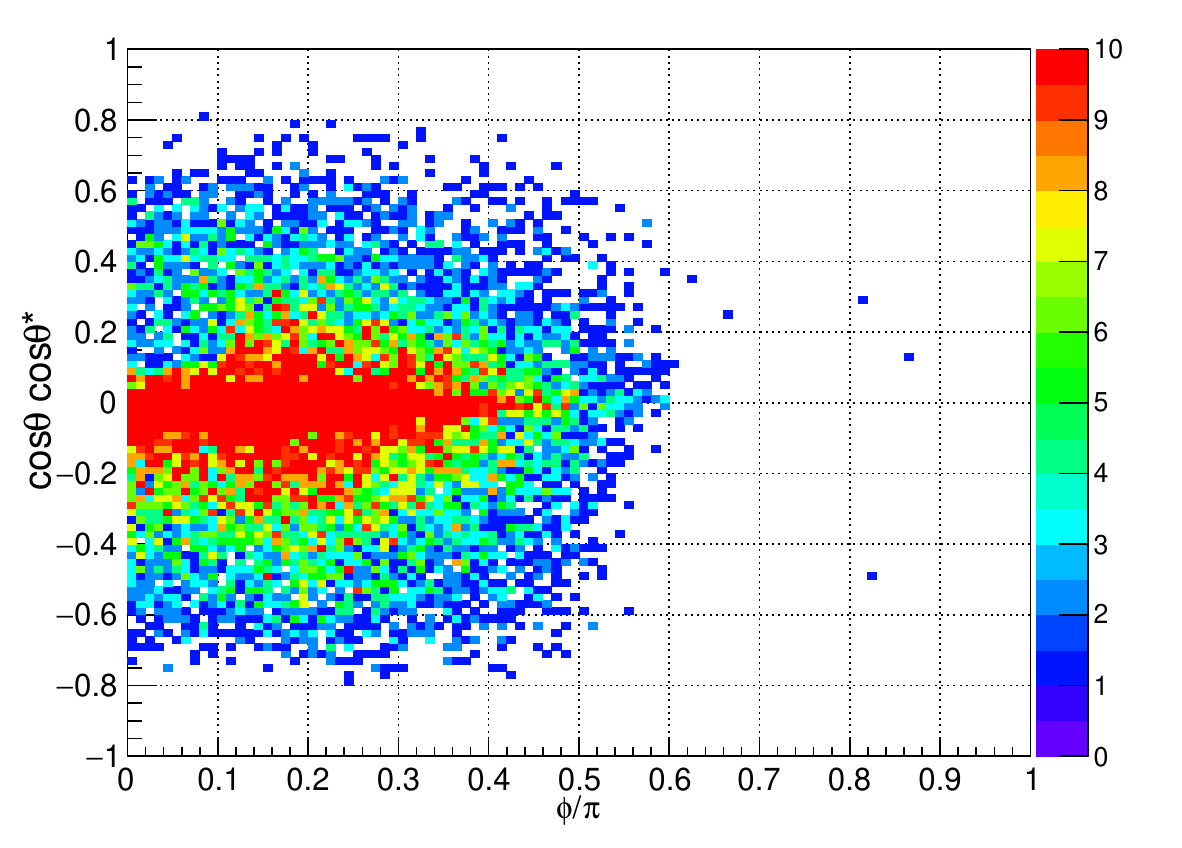}}  
    \subfloat[$\bar {\sigma}_1 (h_{3}^{Z}) <$ 0]{\includegraphics[width=.48\columnwidth]{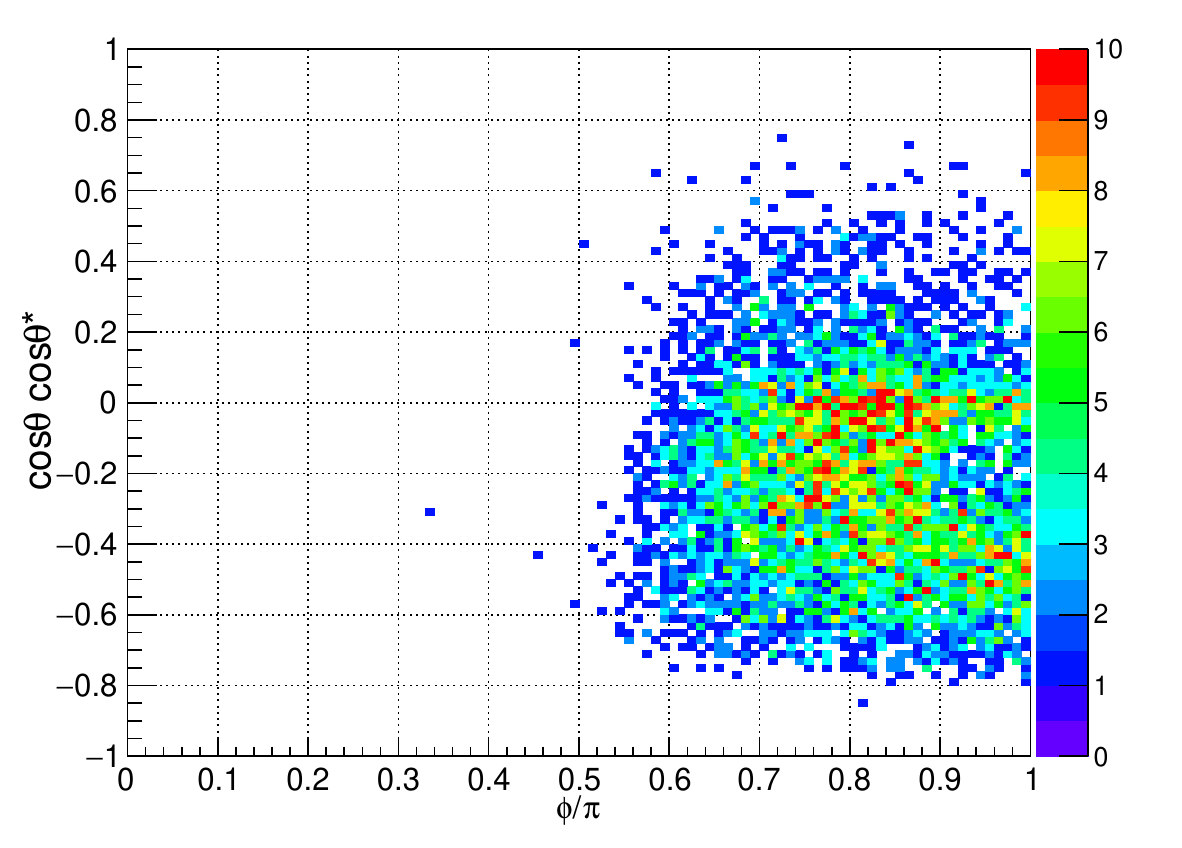}}  \\
\vspace*{-1mm} 
\caption{Normalized distributions of the angular variables employed to analyze the form factors $h_{4}^{}$, $h_{3}^{\gamma}$, and $h_{3}^{Z}$ 
by using simulated interference events.\ The left panel presents events with positive cross section 
in the interference term, whereas the right panel corresponds to events with negative cross section 
for the interference term.}
\label{fig:BDTG_separation}
\end{figure}

Additionally, we use the Toolkit for Multivariate Data Analysis (TMVA)\,\cite{TMVA_toturial,hoecker2009tmva}, a component within the ROOT~\cite{ROOT_toturial} framework for analyzing complex data sets, which provides a broad range of machine learning methods for classification and performance enhancement.\  Also, we employ the Boosted Decision Trees with Gradient boosting (BDTG) algorithm, 
which is a powerful tool for multivariate analysis with a broad range of classification algorithms.\  
Its incorporation helps TMVA to process data more accurately and efficiently, making it a valuable asset for detailed data analysis. 

The 2-dimensional distributions used in our multivariate study leverage measurements of three angles: $\phi$ denotes the angle between the scattering plane and the decay plane of $Z$ boson, $\theta$ is the polar angle of the outgoing $Z$ with respect to the initial electron as introduced in Fig.\,\ref{fig:phi_angle}, and $\theta^*$ is the decay angle 
as measured in the $Z$ boson's rest frame.\ 
The distributions of these angles shown in 
Fig.\,\ref{fig:BDTG_separation} 
are key elements in this study.
In this measurement, we defined a new variable $u = \cos\theta \times \cos\theta^*$.
By employing the BDTG method, we constructed a decision boundary in the $\phi - u$ plane to effectively distinguish between events with positive and negative cross sections.

The normalized 2-dimensional distributions of $\phi$ versus $\cos\theta \cos\theta^*$ demonstrate notable contrasts, indicating the critical importance of interference effects  
for constraining the nTGC form factors and the SMEFT parameters.\ 
The left panels of Fig.\,\ref{fig:BDTG_separation} 
display events with positive interference cross sections, 
whereas the right panels present those with negative values.\ 
These contrasts are instrumental in highlighting the influence of the interference terms 
and facilitating the extraction of the nTGC form factors.

\begin{table}[!h]
    \centering
    \begin{tabular}{c | c c | c c | c}
        \hline
        \hline
                            & \multicolumn{2}{c}{Truth}   & \multicolumn{2}{|c|}{With BDTG}  & \\
        \hline
        Form factor         & Positive & Negative         & Positive & Negative             & Net    \\
        \hline
        $h_{4}$             & 284.01   & -197.89           & 275.93   & -189.82               & 86.12  \\
        $h_{3}^{\gamma}$    & 279.48   & -16.20            & 264.03   & -0.75                 & 263.27 \\
        $h_{3}^{Z}$         & 52.35    & -20.45            & 52.27    & -20.37                & 31.90  \\
        \hline
        \hline
    \end{tabular}
    \caption{Cross sections in fb for the interference terms $\bar\sigma_1$ of $h_{4}$, $h_{3}^{\gamma}$, and $h_{3}^{Z}$ for positive and negative signs, comparing the Monte Carlo truth values and those obtained when the BDTG method is applied.}
    \label{tab:xs_BDTG}
\end{table}

Positive events in the ``Truth" column in Table\,\ref{tab:xs_BDTG} correspond to the distributions (a), (c), and (e) in Fig.\,\ref{fig:BDTG_separation}.
Conversely, negative events denote the distributions (b), (d), and (f). 
The positive and negative events are indistinguishable, necessitating the use of the BDTG method for setting separation boundaries.
The cross sections extracted for both positive and negative events using the BDTG method are presented in the ``With BDTG" column of Table\,\ref{tab:xs_BDTG}.

Using this method, the cross sections for positive and negative events are precisely calculated and applied in the fitting process instead of using the net cross section.\  
The effective cross sections used in the fitting are the sum of the absolute values of positive and negative cross sections.\ 
This approach results in larger effective cross sections compared to the net cross sections, leading to improved sensitivity in the measurements.

%% file: Template.bib
@article{Degrande_2014,
   title={A basis of dimension-eight operators for anomalous neutral triple gauge boson interactions},
   volume={2014},
   number={2},
   journal={JHEP},
   publisher={Springer Science and Business Media LLC},
   author={Degrande, Celine},
   year={2014},
   eprint = {1308.6323},
   archivePrefix = {arXiv},
   primaryClass={hep-ph}
}

@article{Ellis_2023_nny,  
   title={Probing neutral triple gauge couplings with $Z\gamma$ $(\nu\bar{\nu}\gamma$) production at hadron colliders},
   journal={Physical Review D 108 (2023) L111704, no.11,},
   publisher={American Physical Society (APS)},
   author={Ellis, John and He, Hong-Jian and Xiao, Rui-Qing},
   eprint = {2308.16887},
   archivePrefix = {arXiv},
   primaryClass={hep-ph}
}

@article{Ellis_2023_lly,
   title={Probing neutral triple gauge couplings at the LHC and future hadron colliders},
   journal={Physical Review D 107 (2023) 035005, no.3,},
   publisher={American Physical Society (APS)},
   author={Ellis, John and He, Hong-Jian and Xiao, Rui-Qing},
   eprint = {2206.11676},
   archivePrefix = {arXiv},
   primaryClass = {hep-ph}
}

@article{Murphy:2020rsh,
    author = "Murphy, Christopher W.",
    title = "{Dimension-8 operators in the Standard Model Eective Field Theory}",
    eprint = "2005.00059",
    archivePrefix = "arXiv",
    primaryClass = "hep-ph",

    journal = "JHEP",
    volume = "10",
    pages = "174",
    year = "2020"
}

@article{Li:2020gnx,
    author = "Li, Hao-Lin and Ren, Zhe and Shu, Jing and Xiao, Ming-Lei and Yu, Jiang-Hao and Zheng, Yu-Hui",
    title = "{Complete set of dimension-eight operators in the standard model effective field theory}",
    eprint = "2005.00008",
    archivePrefix = "arXiv",
    primaryClass = "hep-ph",
    journal = "Phys. Rev. D",
    volume = "104",
    number = "1",
    pages = "015026",
    year = "2021"
}

@article{Ellis_2020,
   title={Probing new physics in dimension-8 neutral gauge couplings at $e^{+}e^{-}$ colliders},
   journal={Science China (Phys. Mech. \& Astron.) 64 (2020) 221062, no.2,},
   publisher={Springer Science and Business Media LLC},
   author={Ellis, John and He, Hong-Jian and Xiao, Rui-Qing},
   eprint = {2008.04298},
   archivePrefix = {arXiv},
   primaryClass = {hep-ph}
}

@article{NewEFTnTGC,
   title={Probing the scale of new physics in the ZZ$\gamma$ coupling at $e^{+}e^{-}$ colliders},
   journal={Chinese Physics C 44 (2020) 063106, no.6,},
   publisher={IOP Publishing},
   author={Ellis, John and Ge, Shao-Feng and He, Hong-Jian and Xiao, Rui-Qing},
   eprint = {1902.06631},
   archivePrefix = {arXiv},
   primaryClass = {hep-ph}
}

@misc{
CEPCcdr,
      title={CEPC Conceptual Design Report: Volume 2 - Physics $\&$ Detector}, 
      author={The CEPC Study Group},
      year={2018},
      eprint={1811.10545},
      archivePrefix={arXiv},
      primaryClass={hep-ex}
}

@article{
IntroductionMG5,
    author = {Marco Zaro},
    title = {An introduction to MadGraph5\_aMC@NLO},
    year = {2022},
    url = {https://pcteserver.mi.infn.it/~mzaro/mg5amc-tif2-2022/tutorial-unimi-2022-tif.pdf}
}

@article{
IntroductionPY8,
   title={A brief introduction to PYTHIA 8.1},
   volume={178},
   number={11},
   journal={Computer Physics Communications},
   publisher={Elsevier BV},
   author={Sjöstrand, Torbjörn and Mrenna, Stephen and Skands, Peter},
   year={2008},
   eprint = {0710.3820},
   archivePrefix = {arXiv},
   primaryClass = {hep-ph}
}

@article{
IntroductionWhizard, 
   title={WHIZARD—simulating multi-particle processes at LHC and ILC},
   volume={71},
   number={9},
   journal={The European Physical Journal C},
   publisher={Springer Science and Business Media LLC},
   author={Kilian, Wolfgang and Ohl, Thorsten and Reuter, Jürgen},
   year={2011},
   eprint = {0708.4233},
   archivePrefix = {arXiv},
   primaryClass = {hep-ph}
}

@inproceedings{
IntroductionMokka,
    author = {Mora de Freitas, P. and Videau, H.},
    title = {Detector simulation with MOKKA / GEANT4: Present and future},
    booktitle = {International Workshop on Linear Colliders (LCWS 2002)},
    reportNumber = {LC-TOOL-2003-010},
    pages = {623--627},
    month = {8},
    year = {2002},
    eprint = {1902.06161v2},
    archivePrefix = {arXiv},
    primaryClass = {physics.ins-det}
}

@article{
IntroductionG4,
    author = {Agostinelli, S. and others},
    collaboration = {GEANT4},
    title = {GEANT4--a simulation toolkit},
    reportNumber = {SLAC-PUB-9350, FERMILAB-PUB-03-339, CERN-IT-2002-003},
    journal = {Nucl. Instrum. Meth. A},
    volume = {506},
    pages = {250--303},
    year = {2003}
}

@article{
IntroductionDelphes,
    author = {Chen, Cheng and Mo, Xin and Selvaggi, Michele and Li, Qiang and Li, Gang and Ruan, Manqi and Lou, Xinchou},
    title = {Fast simulation of the CEPC detector with Delphes},
    eprint = {1712.09517},
    archivePrefix = {arXiv},
    primaryClass = {hep-ex},
    month = {12},
    year = {2017}
}

@article{
IntroductionPFA_Manqi,
    title={Reconstruction of physics objects at the Circular Electron Positron Collider with Arbor},
    volume={78},
    number={5},
    journal={The European Physical Journal C},
    publisher={Springer Science and Business Media LLC},
    author={Ruan, Manqi and Zhao, Hang and Li, Gang and Fu, Chengdong and Wang, Zhigang and Lou, Xinchou and Yu, Dan and Boudry, Vincent and Videau, Henri and Balagura, Vladislav and Brient, Jean-Claude and Lai, Peizhu and Kuo, Chia-Ming and Liu, Bo and An, Fenfen and Chen, Chunhui and Prell, Soeren and Li, Bo and Laketineh, Imad},
    year={2018},
    eprint = {1806.04879},
    archivePrefix = {arXiv},
    primaryClass = {hep-ex}
}

@article{
IntroductionArbor,
    title={Arbor, a new approach of the Particle Flow Algorithm}, 
    author={Manqi Ruan},
    year={2014},
    eprint={1403.4784},
    archivePrefix={arXiv},
    primaryClass={physics.ins-det}
}

@article{
LeptonID,
   title={Lepton identification at particle flow oriented detector for the future $e^{+}e^{-}$ Higgs factories},
   volume={77},
   number={9},
   journal={The European Physical Journal C},
   publisher={Springer Science and Business Media LLC},
   author={Yu, Dan and Ruan, Manqi and Boudry, Vincent and Videau, Henri},
   year={2017},
   eprint = {1701.07542},
   archivePrefix = {arXiv},
   primaryClass = {physics.ins-det}
}

@article{
CEPCZH_BaiYu,
    title = {Measurements of decay branching fractions of $H\rightarrow b\bar{b}/c\bar{c}/gg$ in associated ($e^{+}e^{-}/\mu^+\mu^-$)H production at the CEPC},
    volume={44},
    number={1},
    journal={Chinese Physics C},
    publisher={IOP Publishing},
    author={Bai, Yu and Chen, Chun-Hui and Fang, Ya-Quan and Li, Gang and Ruan, Man-Qi and Shi, Jing-Yuan and Wang, Bo and Kong, Pan-Yu and Lan, Bo-Yang and Liu, Zhan-Feng},
    year={2020},
    pages={013001},
    eprint = {1905.12903},
    archivePrefix = {arXiv},
    primaryClass = {hep-ex}
}

@article{
CEPCyy_FY,
    author = "Guo, Fangyi and Fang, Yaquan and Li, Gang and Lou, Xinchou",
    title = "{Expected measurement precision of the branching ratio of the Higgs boson decaying to the di-photon at the CEPC}",
    eprint = "2205.13269",
    archivePrefix = "arXiv",
    primaryClass = "hep-ex",
    reportNumber = "Chinese Physics C Vol. 47, No. 4 (2023) 043002",
    journal = "Chin. Phys. C",
    volume = "47",
    number = "4",
    pages = "043002",
    year = "2023"
}

@misc{
EFTfun,
    author = {EFT-fun},
    url    = {https://gitlab.cern.ch/eft-tools/eft-fun}
}

@misc{
TMVA_toturial,
    title = {TMVA, Toolkit for Multivariate Data Analysis with ROOT},
    url = {https://root.cern.ch/download/doc/tmva/TMVAUsersGuide.pdf}, 
}

@misc{
hoecker2009tmva,
      title={TMVA - Toolkit for Multivariate Data Analysis}, 
      author={A. Hoecker and P. Speckmayer and J. Stelzer and J. Therhaag and E. von Toerne and H. Voss and M. Backes and T. Carli and O. Cohen and A. Christov and D. Dannheim and K. Danielowski and S. Henrot-Versille and M. Jachowski and K. Kraszewski and A. Krasznahorkay Jr. au2 and M. Kruk and Y. Mahalalel and R. Ospanov and X. Prudent and A. Robert and D. Schouten and F. Tegenfeldt and A. Voigt and K. Voss and M. Wolter and A. Zemla},
      year={2009},
      eprint={physics/0703039},
      archivePrefix={arXiv},
      primaryClass={physics.data-an}
}

@misc{
ROOT_toturial,
    title = {ROOT : analyzing petabytes of data, scientifically},
    url = {https://root.cern/},
}

@article{Grzadkowski:2010es,
    author = "Grzadkowski, B. and Iskrzynski, M. and Misiak, M. and Rosiek, J.",
    title = "{Dimension-Six Terms in the Standard Model Lagrangian}",
    eprint = "1008.4884",
    archivePrefix = "arXiv",
    primaryClass = "hep-ph",
    reportNumber = "IFT-9-2010, TTP10-35",
    journal = "JHEP",
    volume = "10",
    pages = "085",
    year = "2010"
}

@article{Giudice:2007fh,
    author = "Giudice, G. F. and Grojean, C. and Pomarol, A. and Rattazzi, R.",
    title = "{The Strongly-Interacting Light Higgs}",
    eprint = "hep-ph/0703164",
    archivePrefix = "arXiv",
    reportNumber = "CERN-PH-TH-2007-47",
    journal = "JHEP",
    volume = "06",
    pages = "045",
    year = "2007"
}

@article{Berthier:2015oma,
    author = "Berthier, Laure and Trott, Michael",
    title = "{Towards consistent Electroweak Precision Data constraints in the SMEFT}",
    eprint = "1502.02570",
    archivePrefix = "arXiv",
    primaryClass = "hep-ph",
    journal = "JHEP",
    volume = "05",
    pages = "024",
    year = "2015"
}

@article{Berthier:2015gja,
    author = "Berthier, Laure and Trott, Michael",
    title = "{Consistent constraints on the Standard Model Effective Field Theory}",
    eprint = "1508.05060",
    archivePrefix = "arXiv",
    primaryClass = "hep-ph",
    journal = "JHEP",
    volume = "02",
    pages = "069",
    year = "2016"
}

@article{Biekotter:2018ohn,
    author = {Biek\"otter, Anke and Corbett, Tyler and Plehn, Tilman},
    title = "{The Gauge-Higgs Legacy of the LHC Run II}",
    eprint = "1812.07587",
    archivePrefix = "arXiv",
    primaryClass = "hep-ph",
    journal = "SciPost Phys.",
    volume = "6",
    number = "6",
    pages = "064",
    year = "2019"
}

@article{Ellis:2020unq,
    author = "Ellis, John and Madigan, Maeve and Mimasu, Ken and Sanz, Veronica and You, Tevong",
    title = "{Top, Higgs, Diboson and Electroweak Fit to the Standard Model Effective Field Theory}",
    eprint = "2012.02779",
    archivePrefix = "arXiv",
    primaryClass = "hep-ph",
    reportNumber = "KCL-PH-TH/2020-73, CERN-TH-2020-202",
    journal = "JHEP",
    volume = "04",
    pages = "279",
    year = "2021"
}

@article{Corbett:2021eux,
    author = "Corbett, Tyler and Helset, Andreas and Martin, Adam and Trott, Michael",
    title = "{EWPD in the SMEFT to dimension eight}",
    eprint = "2102.02819",
    archivePrefix = "arXiv",
    primaryClass = "hep-ph",
    journal = "JHEP",
    volume = "06",
    pages = "076",
    year = "2021"
}

@article{Corbett:2023qtg,
    author = "Corbett, Tyler and Desai, Jay and \'Eboli, O. J. P. and Gonzalez-Garcia, M. C. and Martines, Matheus and Reimitz, Peter",
    title = "{Impact of dimension-eight SMEFT operators in the electroweak precision observables and triple gauge couplings analysis in universal SMEFT}",
    eprint = "2304.03305",
    archivePrefix = "arXiv",
    primaryClass = "hep-ph",
    reportNumber = "YITP-SB-2023-04, FERMILAB-PUB-23-134-V, UWThPh 2023-13",
    journal = "Phys. Rev. D",
    volume = "107",
    number = "11",
    pages = "115013",
    year = "2023"
}

@article{Brivio:2019ius,
    author = "Brivio, Ilaria and Bruggisser, Sebastian and Maltoni, Fabio and Moutafis, Rhea and Plehn, Tilman and Vryonidou, Eleni and Westhoff, Susanne and Zhang, C.",
    title = "{O new physics, where art thou? A global search in the top sector}",
    eprint = "1910.03606",
    archivePrefix = "arXiv",
    primaryClass = "hep-ph",
    reportNumber = "P3H-19-036, CERN-TH-2019-193",
    journal = "JHEP",
    volume = "02",
    pages = "131",
    year = "2020"
}

@article{Durieux:2019rbz,
    author = {Durieux, Gauthier and Irles, Adrian and Miralles, V\'\i{}ctor and Pe\~nuelas, Ana and P\"oschl, Roman and Perell\'o, Mart\'\i{}n and Vos, Marcel},
    title = "{The electro-weak couplings of the top and bottom quarks \textemdash{} Global fit and future prospects}",
    eprint = "1907.10619",
    archivePrefix = "arXiv",
    primaryClass = "hep-ph",
    reportNumber = "IFIC/19-33, FIC/19-33, FTUV/19-0724",
    journal = "JHEP",
    volume = "12",
    pages = "98",
    year = "2019",
    note = "[Erratum: JHEP 01, 195 (2021)]"
}

@article{Ethier:2021bye,
    author = "Ethier, Jacob J. and Magni, Giacomo and Maltoni, Fabio and Mantani, Luca and Nocera, Emanuele R. and Rojo, Juan and Slade, Emma and Vryonidou, Eleni and Zhang, Cen",
    collaboration = "SMEFiT",
    title = "{Combined SMEFT interpretation of Higgs, diboson, and top quark data from the LHC}",
    eprint = "2105.00006",
    archivePrefix = "arXiv",
    primaryClass = "hep-ph",
    reportNumber = "OUTP-20-05P, Nikhef-2020-020, CP3-21-12, MCNET-21-07, MAN/HEP/2021/004",
    journal = "JHEP",
    volume = "11",
    pages = "089",
    year = "2021"
}

@article{Buchmuller:1985jz,
    author = "Buchmuller, W. and Wyler, D.",
    title = "{Effective Lagrangian Analysis of New Interactions and Flavor Conservation}",
    reportNumber = "CERN-TH-4254/85",
    journal = "Nucl. Phys. B",
    volume = "268",
    pages = "621--653",
    year = "1986"
}

@article{Pomarol:2013zra,
    author = "Pomarol, Alex and Riva, Francesco",
    title = "{Towards the Ultimate SM Fit to Close in on Higgs Physics}",
    eprint = "1308.2803",
    archivePrefix = "arXiv",
    primaryClass = "hep-ph",
    journal = "JHEP",
    volume = "01",
    pages = "151",
    year = "2014"
}

@article{deBlas:2019rxi,
    author = "de Blas, J. and others",
    title = "{Higgs Boson Studies at Future Particle Colliders}",
    eprint = "1905.03764",
    archivePrefix = "arXiv",
    primaryClass = "hep-ph",
    reportNumber = "DESY-19-079",
    journal = "JHEP",
    volume = "01",
    pages = "139",
    year = "2020"
}

@article{He:2015spf,
    author = "He, Hong-Jian and Ren, Jing and Yao, Weiming",
    title = "{Probing new physics of cubic Higgs boson interaction via Higgs pair production at hadron colliders}",
    eprint = "1506.03302",
    archivePrefix = "arXiv",
    primaryClass = "hep-ph",
    journal = "Phys. Rev. D",
    volume = "93",
    number = "1",
    pages = "015003",
    year = "2016"
}

@article{Ge:2016zro,
    author = "Ge, Shao-Feng and He, Hong-Jian and Xiao, Rui-Qing",
    title = "{Probing new physics scales from Higgs and electroweak observables at $e^{+}e^{-}$ Higgs factory}",
    eprint = "1603.03385",
    archivePrefix = "arXiv",
    primaryClass = "hep-ph",
    journal = "JHEP 10 (2016) 007",
}

@article{Ge:2016tmm,
    author = "Ge, Shao-Feng and He, Hong-Jian and Xiao, Rui-Qing",
    title = "{Testing Higgs coupling precision and new physics scales at lepton colliders}",
    eprint = "1612.02718",
    archivePrefix = "arXiv",
    primaryClass = "hep-ph",
    journal = "Int. J. Mod. Phys. A 31 (2016) 1644004, no.33",
}

@article{Jahedi:2022duc,
    author = "Jahedi, Sahabub and Lahiri, Jayita",
    title = "{Probing anomalous ZZ\ensuremath{\gamma} and Z\ensuremath{\gamma}\ensuremath{\gamma} couplings at the $e^+e^-$ colliders using optimal observable technique}",
    eprint = "2212.05121",
    archivePrefix = "arXiv",
    primaryClass = "hep-ph",
    doi = "10.1007/JHEP04(2023)085",
    journal = "JHEP",
    volume = "04",
    pages = "085",
    year = "2023"
}

@article{Jahedi:2023myu,
    author = "Jahedi, Sahabub",
    title = "{Optimal estimation of dimension-8 neutral triple gauge couplings at the $e^+e^-$ colliders}",
    eprint = "2305.11266",
    archivePrefix = "arXiv",
    primaryClass = "hep-ph",
    doi = "10.1007/JHEP12(2023)031",
    journal = "JHEP",
    volume = "12",
    pages = "031",
    year = "2023"
}

@article{Zhang:2024bld,
    author = "Zhang, Yulei and Mo, Cen and Chen, Xiang and Li, Bingzhi and Chen, Hongyang and Hu, Jifeng and Li, Liang",
    title = "{Search for Long-lived Particles at Future Lepton Colliders Using Deep Learning Techniques}",
    eprint = "2401.05094",
    archivePrefix = "arXiv",
    primaryClass = "hep-ex",
    month = "1",
    year = "2024"
}

@article{CEPCStudyGroup:2023quu,
    author = "Abdallah, Waleed and others",
    collaboration = "CEPC Study Group",
    title = "{CEPC Technical Design Report -- Accelerator (v2)}",
    eprint = "2312.14363",
    archivePrefix = "arXiv",
    primaryClass = "physics.acc-ph",
    reportNumber = "IHEP-CEPC-DR-2023-01, IHEP-AC-2023-01",
    month = "12",
    year = "2023"
}
